# Significant Role of DNA Backbone in Mediating the Transition Origin of Electronic Excitations of B-DNA – Implication from Long Range Corrected TDDFT and Quantified NTO Analysis

Jian-Hao Li[a, b], Jeng-Da Chai[a, c, *], Guang-Yu Guo[a, c, d], Michitoshi Hayashi[b, *]

[a]*Department of Physics, Center for Theoretical Sciences, National Taiwan University, Taipei 10617, Taiwan.*
[b]*Center for Condensed Matter Sciences, National Taiwan University, Taipei 10617, Taiwan.*
[c]*Center for Quantum Science and Engineering, National Taiwan University, Taipei 10617, Taiwan.*
[d]*Graduate Institute of Applied Physics, National Chengchi University, Taipei 11605, Taiwan*
E-mail: jdchai@phys.ntu.edu.tw (J.-D. Chai); atmyh@ntu.edu.tw (M. Hayashi)
2nd November 2011

We systematically investigate the possible complex transition origin of electronic excitations of giant molecular systems by using the recently proposed QNTO analysis [J.-H. Li, J.-D. Chai, G. Y. Guo and M. Hayashi, *Chem. Phys. Lett.*, 2011, **514**, 362.] combined with long-range corrected TDDFT calculations. Thymine (Thy) related excitations of biomolecule B-DNA are then studied as examples, where the model systems have been constructed extracting from the perfect or a X-ray crystal (PDB code 3BSE) B-DNA structure with at least one Thy included. In the first part, we consider the systems composed of a core molecular segment (e.g. Thy, di-Thy) and a surrounding physical/chemical environment of interest (e.g. backbone, adjacent stacking nucleobases) and examine how the excitation properties of the core vary in response to the environment. We find that the orbitals contributed from DNA backbone and surrounding nucleobases often participate in a transition of Thy-related excitations affecting their composition, absorption energy, and oscillator strength. In the second part, we take into account geometrically induced variation of the excitation properties of various B-DNA segments, e.g. di-Thy, dTpdT etc., obtained from different sources (ideal and 3BSE). It is found that the transition origin of several Thy-related excitations of these segments is sensitive to slight conformational variations, suggesting that DNA with thermal motions in cells may from time to time exhibit very different photo-induced physical and/or chemical processes.

## 1. Introduction

The primary functions of DNA molecules to encode cellular genetic information and to engage in replication have readily rendered them to be the most important biomolecules[1, 2]. Investigation on them has long been a hot topic to date such that a great deal of understanding has been advanced on their mechanisms, physical properties, chemical properties etc. and possible application in nanotechnology[3-6]. Among all the important issues on DNA, one is related to their optical absorption spectroscopy. It has been well-known that the optical absorption spectra of nucleobases – thymine, adenine, cytosine, guanine – or different combinations of them are dominated by bright $^1\pi\pi^*$ excitations interlaced with dark $^1n\pi^*$ ones in the ultraviolet range[7-35]. Its related experimental and theoretical studies can be reviewed, for example, in an excellent collection[36]; several quantum chemistry methods, like configuration interaction singles (CIS)[37-43], complete active space self-consistent-field with multiconfigurational second order perturbation theory (CASPT2/CASSCF)[44-46], second-order approximate coupled-cluster theory with the resolution-of-the-identity approximation (RI-CC2)[47-48], time-dependent DFT (TDDFT)[49-57] or etc. have been used in literature.

In order to know more about how nucleobases alter their excitation properties in response to surrounding environment within a giant complex DNA molecule, however, it is still a formidable task nowadays employing high-level quantum chemistry approaches, like equation of motion coupled cluster singles and doubles (EOM-CCSD)[58-60] and etc., to directly simulate a huge system composed of the core part and the interesting environment. TDDFT[61-64], on the other hand, may currently be the only choice that can deal with large size molecular systems at the same time with less computational difficulty, although the correctness of output can be a debate.

In practice, model potentials and/or embedded system schemes are often employed to divide the whole system into many sub-layers/sub-systems handled by separate theoretical treatments[65-80]. In this way, interesting local excitations of a core subsystem embedded in an environment can be sophisticatedly described. However, a noticeable drawback is that orbital transitions from/to environments are not allowed. Fortunately, a direct simulation of giant molecular systems by (TD-)DFT has recently become feasible and reliable due to the great progress on the development of long-range corrected (LC) hybrid functionals[81-94]. In this case, effects of such orbital transitions can be feasibly included.

In the LC hybrid scheme, the Coulomb operator is first split into a long-range (LR) operator $L(r_{12})/r_{12}$ and a complementary short-range (SR) operator $[1-L(r_{12})]/r_{12}$,

$$\frac{1}{r_{12}} = \frac{L(r_{12})}{r_{12}} + \frac{[1-L(r_{12})]}{r_{12}} \quad (1)$$

where $L(r_{12})$ is a function of inter-electronic separation $r_{12} \equiv |\mathbf{r}_{12}| = |\mathbf{r}_1 - \mathbf{r}_2|$, ranging from 0 to 1, and approaching to 1 at large $r_{12}$. Currently, the most popular $L(r_{12})$ used in the LC hybrid scheme is the standard error function (erf), although there are several variants as well.

After the splitting, the LR part of exchange is treated exactly by Hartree-Fock (HF) theory (where $\gamma_\sigma(\mathbf{r}_1, \mathbf{r}_2)$ is the one-electron spin density matrix),

$$E_x^{LR-HF} = -\frac{1}{2} \sum_\sigma \iint L(r_{12}) \frac{|\gamma_\sigma(\mathbf{r}_1, \mathbf{r}_2)|^2}{r_{12}} d\mathbf{r}_1 d\mathbf{r}_2 \quad (2)$$

the SR part of exchange is treated by density functional approxi-mations (DFAs) (e.g. LDA or GGAs), and the correlation (also treated by DFAs) remains the same as that of the full Coulomb interaction,

$$E_{xc}^{LC-DFA} = E_x^{LR-HF} + E_x^{SR-DFA} + E_c^{DFA} \qquad (3)$$

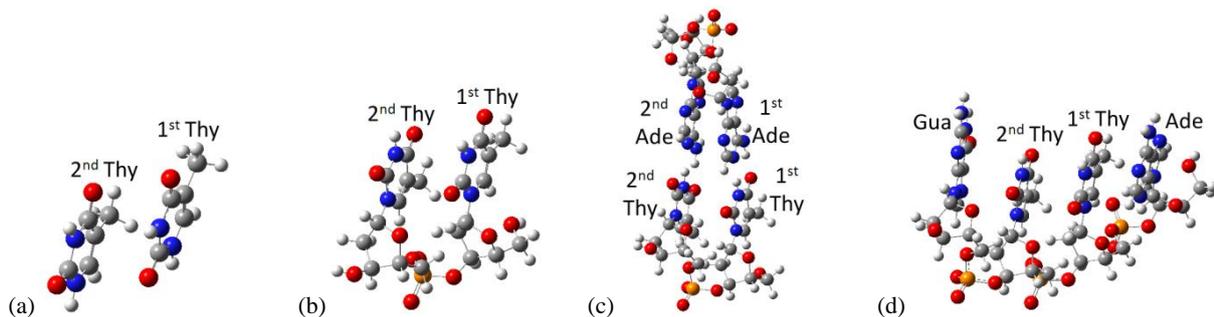

**Fig. 1** Various Thy-comprised systems extracted from ideal B-DNA: (a) two thymines (di-Thy), (b) two thymines connected by DNA backbone (dTpdT), (c) dTpdT with WC pairing nucleobases (dTpdT--dApdA), and (f) dTpdT with adjacent stacking nucleobases (dApdTpdTpdG). The order of bases is counted from 5'-end to 3'-end of the DNA strand.

In recent years, LC hybrid functionals have gained increasing attentions due to their success in several important applications[83-86].

In addition, we have lately proposed quantified natural transition orbital (QNTO) analysis[95] which is able to clearly exhibit the transition origin of electronic excitations. This framework should be especially useful for treating large biomolecules with high orbital energy density, where even the low-lying excitations are generally composed of several singly excited configurations (SECs) with similar weights pairing different occupied and unoccupied orbitals in the linear combination of SECs (LCSEC). This situation may be more commonplace particularly for DNA systems that consist of many similar molecular segments. Under such circumstances, QNTO analysis can often lead to one or two dominant SECs corresponding to NTO1 and NTO2 in the NTO-based LCSEC. Here NTO1(2) stands for the first (second) dominant NTO pair. The NTO1(2) can then be interpreted in terms of a standard-orbitals set chosen depending on the type of systems or chemical interest; for example, if any important photo-induced chemical reaction takes place in a dimer system, orbitals from monomer segment can be chosen as standard-orbitals. In other words, the transition origin of an electronic excitation of dimer can be interpreted by the monomer orbitals.

Thus, in light of the above two developments, the present work is in an effort by using TDDFT with LC hybrid functionals (LC-TDDFT) and QNTO analysis to systematically investigate the electronic excitations of giant molecular systems, here the DNA, in which orbitals exchange between core-subsystem and environment can be important.

We choose a Thy-appearing region of B-DNA (the commonest cellular form of DNA)[96] as an example and focus on its Thy-related excitations in which hole- and electron-orbitals are both significantly contributed from Thy. We study how different surrounding environments can play a role to its transition origin, absorption energy and oscillator strength. This is accomplished by extracting various B-DNA segments (nucleobases or oligonucleotides) with different environmental factors existing in vicinity of a Thy and by looking into their respective electronic excitation properties. Moreover, the excitation properties of B-DNA segments obtained from different sources (ideal and 3BSE) are compared with each other to investigate conformational effect as well.

Recently, Kozak et al. have reported a thorough study[97] on the singlet electronic excitations of two π-stacked Thys through implementing a variety of quantum chemistry methods, including CIS, CIS(2), EOM-CCSD, TDDFT and CASSCF, based on molecular exciton theory[98]. It is found that the coupling of Thy monomer orbitals to form dimer orbitals strongly depend on the distance and the relative orientation of two Thys. However, we have shown by using QNTO analysis[95] that for a real dimer system with low or no symmetry, such as two π-stacked Thys extracted from B-DNA, two local orbital transitions can no longer equally contribute to an electronic excitation predicted by the molecular exciton theory. If the dimer separation is short enough, the additional charge transfer (CT) contribution can also exist. Under such circumstances, QNTO can be a powerful tool to study the transition origins of electronic excitations that may consist of local or CT transition components in such a low-symmetry situation which is commonplace in median and large bio-molecular systems. Lange and Herbert[99] have also applied TDDFT with LC hybrid functionals to obtain the accurate absorption energies of interbase CT electronic excitations that are underestimated by TDDFT calculation with pure (e.g. LDA) or global hybrid (e.g. B3LYP) density functionals. Since both of the two reports focus on bare nucleobases model systems, a remaining question of interest is indeed whether the environmental factors like backbone to what extent affect the electronic excitations of nucleobases.

The paper is organized as follows. In section 2, computational procedures are introduced, where the preparation of various B-DNA sliced systems and DFT/TDDFT calculation details are presented. In subsection 3.1, the interpretation of the excitation properties of various systems calculated by TDDFT with QNTO analysis are introduced. Subsections 3.2~3.6 compare the results of various systems such that the influence of different environmental factors, i.e. the presence of another Thy, backbone, Watson Crick (WC) pairing nucleobases and stacking nucleobases, and conformational variation to the Thy-related excitations is successively focused on. Subsection 3.7 demonstrates the overall backbone influence to several low-lying excitations (not particularly Thy-related ones) of the systems consisting of backbone structure. Conclusion is given in Sec. 4.

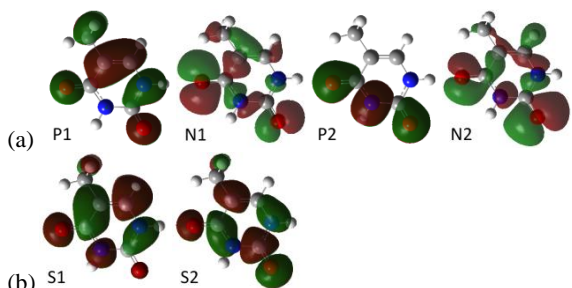

(a) P1  N1  P2  N2
(b) S1  S2

**Fig. 2** Standard-orbitals of Thy generated in the B3LYP-DFT calculation. Isovalue=0.02 is adopted for the graphical representation. P1, N1, P2 and N2 correspond to the HOMO, HOMO-1, HOMO-2, and HOMO-3, respectively, and S1 and S2 are the LUMO and LUMO+1. The P1 and P2 are from the components of $^1\pi\pi^*$-character excitations while N1 and N2 those from $^1n\pi^*$-character excitations.

## 2. Computational Details
### 2.1. Preparation and TDDFT Calculations of Thy-comprised Molecules

Various molecular systems including at least one Thy nucleobase have been prepared by extracting from ideal[100] or X-ray determined (1.6 Å resolution; PDB code 3BSE)[101] B-DNA crystal structure and by saturating each broken chemical bond with a hydrogen atom. Geometry optimization is not performed for these simplified systems to avoid generating unattainable structure of normal DNA motions and no more solvent model is used. The prepared systems are shown in Figures 1 and S1. We investigate (1) environmental effect and (2) conformational effect; for (1) we study how the different environmental surroundings from B-DNA itself affect the Thy-related excitations of a core segment; for (2) we examine the variation of excitation properties of Thy-related excitations resulted from the geometry variation of a segment. For comparison, structures of dTpdT extracted from ideal and 3BSE source are also superimposed together in Figure S2.

Once these molecular systems are built, DFT/TDDFT calculations are carried out using Gaussian09[102] with an economic basis set 6-31G(d)[103] since up to 100 excitations of a system as large as dApdTpdTpdG will be computed in this work. However, the basis effect is also examined where 6-311+G(df,p)[104-105] is tested, shown in Table S3, for a particular system – dTpdT. Moreover, since we focus on the solute form of the B-DNA molecule whose structure is stabilized ionizing one proton for each phosphate, extra electrons are added to each sliced B-DNA segment to account for the charged phosphates. For instance, dTpdT molecule, consisting of one phosphate, is set to have net charge –e. Therefore each molecule is a closed-shell electronic system and we pay attention to the singlet electronic excitations. Various exchange-correlation functionals – PBE1PBE[106], B3LYP[107-108], CAM-B3LYP[82], ωB97X[84], ωB97X-D[83] and LC-ωPBE[81] – are implemented in the DFT/TDDFT calculations. After a TDDFT calculation, QNTO analysis is followed, whose code has been developed locally allowing us to process TDDFT output results generated by Gaussian09.

In the previous report[95] we have found TD-ωB97X among several tested theoretical methods most closely reproduce the transition origin of several low-lying excitations that EOM-CCSD predicts for Thy and dT. Hence, dealing with larger DNA-sliced systems here, we focus on TD-ωB97X results; the results of TDDFT with other functionals are shown in Table S2.

## 3. Results and Discussion
### 3.1. The Excitation Energy, Oscillator Strength, and Transition Origin of Several Low-Lying Thy-related Excitations of Ideal di-Thy, 3BSE dT, Ideal/3BSE dTpdT--dApdA and Ideal/3BSE dApdTpdTpdG

The detailed information of calculated absorption wavelength, oscillator strength and NTO1(2) transition origin of the first 10 selected Thy-related excitations of various systems are listed in Tables 1 and 2. Since we are interested in the influence of surrounding environment and conformational variation to Thy-related excitations, the standard-orbitals are chosen from the monomer orbitals of the ideal/3BSE Thy generated in its B3LYP-DFT calculations. The standard orbitals set can also be generated by other theoretical methods.

Incidentally, the orthogonality of DFT/B3LYP standard orbitals from different moiety of a DNA segment is maintained (mutual projection is below the accuracy, $10^{-2}$, of the projection coefficients) for all the studied cases here. For cases where orthogonality is not fulfilled, such as the two moieties are very close to each other, the standard orbitals cannot be regarded as purely the monomer orbitals and orthogonalization has to be performed first.

Moreover, the targeted core – Thy – is rather intact within different situations studied here so that the Thy stays in the same chemical species, not other isomers with the same composition of atoms. Therefore, its ground state electronic structure remains the same. If the geometry of the target, Thy, is distorted strongly such that it, when let free, will be stabilized to other isomers, its ground state electronic structure is then changed, and the comparison of transition origins of two different species becomes meaningless. In such a case, the same set of standard-orbitals for electron from unoccupied orbitals and for hole from occupied orbitals also cannot be well defined, since some of the occupied and unoccupied orbitals can be switched. In Tables 1 and 2, the selected Thy-related excitations, denoted as SO-hosted (standard-orbitals hosted) excitations thereafter, are those with their hole- (NTO1-H) and electron-orbital (NTO1-E) of NTO1 having overall more than 30% density contribution from the standard-orbitals for hole and for electron, respectively. This is confirmed by first projecting the NTO1(2)-H(E) to the standard-orbitals and the density contribution of a standard-orbital is calculated by its square of projection coefficient. The panels a and b in Figure 2 show the standard-orbitals of the 2nd ideal Thy for hole and for electron, respectively. The other standard-orbitals of the 1st ideal, 1st 3BSE and 2nd 3BSE Thy for hole and for electron are similar to those in Figure 2. Therefore, for example, if the studied core molecular segment is the ideal di-Thy, the standard-orbitals used for projection of a NTO1(2)-H(E) will be the 6 orbitals of the 2nd ideal Thy shown in Figure 2 plus the other 6 similar orbitals from the 1st ideal Thy.

The detailed projection coefficients of the NTO1(2)-H(E) of each SO-hosted excitation are listed in Table S4. Moreover, the transition expression of a NTO1(2) is determined according to

**Table 1** Various excitation properties of (a) ideal Thy, dT (adopted from Ref. 95), (b) ideal di-Thy, dTpdT, dTpdT--dApdA, and dApdTpdTpdG provided by TD-ωB97X calculation and QNTO analysis. N denotes excitation order, while P denotes NTO2 phase for cases where NTO1 has less than 70% domination to the whole excitation. λ (nm) and f stand for absorption wavelength and oscillator strength, respectively. NTO1(2) and % record the expression of transition origin of the first (second) NTO pair and its domination to the whole excitation. Type denotes the excitation classification based on the EOM-CCSD Thy NTO1 expressions used in Ref. 95. In (b) the excitation of a system is referenced to a similar excitation of the other ideal system shown in the subtitle (referred to as the core molecular unit). For instance, the 3rd excitation (Type-A2) of dTpdT is referenced to the 2nd excitation of di-Thy, while the 9th excitation (Type-E2B21) of dTpdT, having no similar type of excitation in di-Thy, is not referenced to any excitation. There is also data – Diff (%) of Energy, Diff(%) of f, and $\sigma_E$ – recording the variation of energy, oscillator strength, and transition origin compared to the result of the referenced system. The local Type-A, B and C derived excitations are marked in bold-face fond in the N column.

(a)

| | Ideal Thy | | | | | | Ideal dT | | | | |
|---|---|---|---|---|---|---|---|---|---|---|---|
| N | λ (nm) | f | NTO1 | % | Type | N | λ (nm) | f | NTO1 | % | Type |
| 1 | 242.13 | 0.0002 | N1t2 - S1t2 | 100 | A2 | 1 | 239.80 | 0.0000 | N1t2 - S1t2 (S2t2) | 99 | (A2) |
| 2 | 228.82 | 0.2013 | P1t2 - S1t2 | 97 | B2 | 2 | 233.96 | 0.2820 | P1t2 - S1t2 | 97 | B2 |
| 3 | 190.78 | 0.0000 | N1t2 N2t2 - (S1t2) S2t2 | 99 | C2 | 3 | 190.28 | 0.0003 | N1t2 N2t2 - (S1t2) S2t2 | 98 | C2 |
| 4 | 178.49 | 0.0659 | P2t2 - S1t2 | 99 | D2 | 4 | 179.36 | 0.1967 | P1t2 - S2t2 | 96 | E2 |
| 5 | 174.43 | 0.1980 | P1t2 - S2t2 | 98 | E2 | 5 | 176.86 | 0.0825 | P2t2 - S1t2 | 97 | D2 |
| 6 | 164.14 | 0.0002 | N2t2 - S1t2 | 97 | F2 | 6 | 168.19 | 0.0015 | N2t2 B (O) - S1t2 | 96 | (F2) |
| 7 | 158.49 | 0.0003 | N1t2 N2t2 - (S1t2) S2t2 | 91 | G2 | 7 | 159.11 | 0.0010 | N1t2 N2t2 - (S1t2) S2t2 | 83 | (G2) |

(b)

| N | λ (nm) ; Diff. (%) of Energy | f ; Diff. (%) of f | NTO1 | $\sigma_E$ | % | Type | N; P | λ (nm) ; Diff. (%) of Energy | f ; Diff. (%) of f | NTO1(2) | $\sigma_E$ | % | Type |
|---|---|---|---|---|---|---|---|---|---|---|---|---|---|
| | Ideal di-Thy (ref. to ideal Thy) | | | | | | | Ideal dTpdT (ref. to ideal di-Thy) | | | | | |
| 1 | 244.16 | -1 | 0.0001 | - | N1t1 - S1t1 | 0.017 | 99 | A1 | 1 | 239.30 | -3 | 0.0370 | -70 | P1t2 - S1t2 | 0.072 | 67 | B2 |
| 2 | 242.42 | 0 | 0.0001 | - | N1t2 - S1t2 | 0.020 | 99 | A2 | - | | | P1t1 - S1t1 | | 30 | |
| 3 | 232.87 | -2 | 0.1254 | -38 | P1t2 - S1t2 | 0.022 | 94 | B2 | 2 | 236.23 | -3 | 0.4157 | 90 | P1t1 - S1t1 | 0.070 | 67 | B1 |
| 4 | 228.92 | 0 | 0.2189 | 9 | P1t1 - S1t1 | 0.044 | 94 | B1 | + | | | P1t2 - S1t2 | | 30 | |
| 5 | 191.44 | 0 | 0 | - | N1t2 N2t2 - (S1t2) S2t2 | 0.041 | 99 | C2 | 3 | 235.97 | 3 | 0.0044 | - | N1t2 - S1t2 (S2t2) | 0.020 | 98 | (A2) |
| 6 | 190.81 | 0 | 0.0011 | - | N1t1 N2t1 - (S1t1) S2t1 | 0.113 | 97 | C1 | 4 | 235.06 | 4 | 0.0001 | - | N1t1 - S1t1 (S2t1) | 0.012 | 98 | (A1) |
| 7 | 188.49 | - | 0.0170 | - | P1t1 - S1t2 (S2t1) | - | 96 | (E1B12) | 6 | 192.01 | -1 | 0.0001 | - | N1t1 N2t1 - (S1t1) S2t1 | 0.030 | 96 | C1 |
| 8 | 181.05 | -1 | 0.0263 | -60 | P2t2 - S1t2 | 0.025 | 91 | D2 | 7 | 189.63 | -1 | 0.0292 | 72 | P1t1 - S1t2 S2t1 | 0.049 | 94 | E1B12 |
| 9 | 179.55 | - | 0.0568 | - | P1t2 P2t1 - S1t1 | - | 84 | D1B21 | 8 | 188.34 | 2 | 0.0037 | - | N1t2 N2t2 - (S1t2) S2t2 | 0.044 | 95 | C2 |
| 10 | 177.84 | - | 0.0749 | - | P1t2 P2t1 - S1t1 (S2t2) | - | 87 | (D2E2B21) | 9 | 186.74 | - | 0.0478 | - | P1t2 B - S1t1 (S2t2) | - | 88 | (E2B21) |
| | | | | | | | | 10 | 185.94 | - | 0.0382 | - | P1t2 B - S1t1 | - | 87 | (B21) |
| | | | | | | | | 12 | 179.79 | - | 0.0912 | - | P1t2 - S1t1 S2t2 | - | 92 | E2B21 |
| | Ideal dTpdT--dApdA (ref. to ideal dTpdT) | | | | | | | Ideal dApdTpdTpdG (ref. to ideal dTpdT) | | | | | |
| 1 | 242.05 | -1 | 0.0383 | 4 | P1t2 - S1t2 | 0.041 | 72 | B2 | 1 | 242.97 | -3 | 0.0480 | -88 | P1t1 - S1t1 | 0.106 | 78 | B1 |
| 2 | 239.09 | -1 | 0.4026 | -3 | P1t1 - S1t1 | 0.046 | 71 | B1 | 2 | 239.62 | 0 | 0.2325 | 528 | P1t2 - S1t2 | 0.100 | 76 | B2 |
| 3 | 224.02 | 5 | 0.0001 | - | N1t1 - S1t1 (S2t1) | 0.018 | 96 | (A1) | 4 | 231.11 | 2 | 0.0001 | - | N1t2 - S1t2 (S2t2) | 0.024 | 94 | (A2) |
| 4 | 223.27 | 6 | 0.0001 | - | N1t2 - S1t2 (S2t2) | 0.029 | 96 | (A2) | 5 | 230.97 | 2 | 0.0001 | - | N1t1 - S1t1 (S2t1) | 0.010 | 93 | (A1) |
| 11 | 193.00 | -1 | 0 | - | (N1t1) N2t1 - (S1t1) S2t1 | 0.024 | 92 | (C1) | 15 | 194.60 | -4 | 0.0363 | -24 | P1t2 - S1t1 (S2t2) | 0.072 | 96 | (E2B21) |
| 16 | 189.76 | -1 | 0.0001 | - | (N1t2) N2t2 - (S1t2) S2t2 | 0.031 | 71 | (C2) | 16 | 192.01 | - | 0.0006 | - | (N2t1) B - S1t1 (S2t1) | - | 67 | (F1) |
| 17 | 188.97 | -2 | 0.0209 | -45 | P1t2 - S1t1 | 0.095 | 92 | (B21) | - | | | (N1t1) N2t1 B - S1t1) S2t1 | | 28 | |
| 19 | 186.79 | 2 | 0.0586 | 101 | P1t1 - S1t2 (S2t1) | 0.039 | 90 | (E1B12) | 18 | 190.91 | 1 | 0.0004 | - | (N1t1) N2t1 B - S1t1 S2t1 | 0.163 | 59 | (C1) |
| 25 | 181.87 | - | 0.0042 | - | P2t1 B - S1t1 | - | 83 | (D1) | + | | | (N1t1) (N2t1) B - S1t1 S2t1 | | 34 | |
| 26 | 180.96 | - | 0.0050 | - | P2t2 A2 - S1t2 | - | 81 | (D2) | 21 | 188.45 | 1 | 0.0562 | 92 | P1t1 - (S1t2) S2t1 (A) | 0.122 | 80 | (E1B12) |
| | | | | | | | | 24 | 186.62 | 1 | 0 | - | N1t2 N2t2 - (S1t2) S2t2 | 0.031 | 93 | C2 |
| | | | | | | | | 28 | 182.31 | - | 0.0940 | - | P1t2 - S2t2 (G) | - | 68 | (E2) |
| | | | | | | | | + | | | P1t1 B - S1t2 (A) (G) | | 14 | |

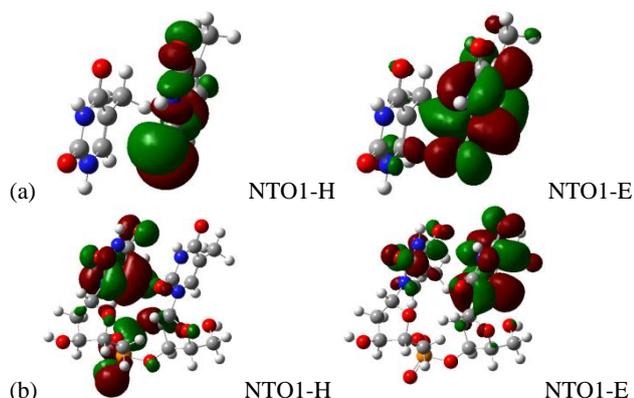

(a) NTO1-H    NTO1-E

(b) NTO1-H    NTO1-E

**Fig. 3** Two examples of SO-hosted excitations of ideal di-Thy and dTpdT molecules. (a) The NTO1-H and NTO1-E plot for the 6$^{th}$ excitation (190.81nm, f=0.0011) of ideal di-Thy. NTO1 accounts for 97% of the whole excitation. With QNTO analysis the transition origin can be expressed in terms of standard-orbitals as "N1t1 N2t1 - (S1t$\underline{1}$) S2t1". (b) The NTO1 plot for the 9$^{th}$ excitation (186.74nm, f=0.0478) of ideal dTpdT. It accounts for 88% of the whole excitation and the transition origin reads "P1t$\underline{2}$ B - S1t1 (S2t2)".

these coefficients. To denote the transition expression, we introduce several notations: for example, P1t2 refers to P1 of the 2nd Thy with a positive coefficient in a system containing two Thys, e.g. di-Thy, dTpdT, while P1t$\underline{2}$ corresponds to P1 with a negative coefficient. P1t12 is an abbreviation of P1t1 plus P1t2 which both has density contribution greater than 0.3 in a NTO1(2)-H. Similarly, P1t1$\underline{2}$ corresponds to the case having combination of P1t1 and P1t$\underline{2}$. In addition, the parenthesis "( )" in P1t(1)2 indicates that it is made up of P1t1 and P1t2 but the coefficient of P1t1 is $(0.3, 0.1]$.

The integration of the orbitals originated from backbone structure contributing to a NTO1(2)-H is denoted as B in a transition expression, whereas the integration of Thy orbitals other than the ones shown in the panels a and b in Figure 2 for NTO1(2)-H and for NTO1(2)-E, respectively, are both denoted as O. A1 and A2 are also likewise introduced for dTpdT--dApdA in a transition expression indicating the integrated contribution of orbitals from the 1st Ade and 2nd Ade, while A and G are introduced for dApdTpdTpdG indicating orbital-contribution from Ade and Gua. If the contribution of the orbital-fraction (B, O, A1, A2, A or G) is $(0.3, 0.1]$, it is shown in "( )" in the transi-tion expression.

**Table 2** Various excitation properties of the first 10 (8 for dT) SO-hosted singlet excitations of X-ray determined (3BSE) dT, dTpdT, dTpdT--dApdA and dApdTpdTpdG provided by TD-ωB97X calculation and QNTO analysis.

| N; P | λ (nm) ; Diff. (%) of Energy | f ; Diff. (%) of f | NTO1(2) | $\sigma_E$ | % | Type | N; P | λ (nm) ; Diff. (%) of Energy | f ; Diff. (%) of f | NTO1(2) | $\sigma_E$ | % | Type |
|---|---|---|---|---|---|---|---|---|---|---|---|---|---|
| | 3BSE dT (ref. to ideal dT) | | | | | | | 3BSE dTpdT (ref. to ideal dTpdT) | | | | | |
| 1 | 238.04 ; 1 | 0.0005 ; - | N1 - S1 (S2) | 0.018 | 99 | (A) | 1 | 238.34 ; -1 | 0.0001 ; - | N1t1 - S1t1 | 0.016 | 98 | (A1) |
| 2 | 228.90 ; 2 | 0.2880 ; 2 | P1 - S1 | 0.016 | 97 | B | 2 | 235.85 ; 0 | 0.0330 ; -92 | P1t1($\underline{2}$) - S1t1($\underline{2}$) | 0.221 | 56 | (B1) |
| 3 | 191.10 ; 0 | 0.0015 ; - | N1 N2 - ($\underline{S1}$) S2 | 0.026 | 98 | C | - | | | P1t(1)2 (N1t2) - S1t(1)2 | | 40 | |
| 4 | 176.81 ; 0 | 0.0899 ; 9 | P2 - S1 | 0.021 | 91 | D | 3 | 234.90 ; 0 | 0.0121 ; - | N1t2 - S1t2 | 0.023 | 87 | (A2) |
| 5 | 176.06 ; 2 | 0.1881 ; -4 | P1 - S2 | 0.021 | 89 | E | 4 | 231.25 ; 3 | 0.4125 ; 1015 | P1t$\underline{2}$ - S1t2 | 0.126 | 64 | B2 |
| 6 | 168.14 ; 0 | 0.0004 ; - | N2 B - S1 | 0.024 | 96 | (F) | + | | | P1t1 - S1t1 | | 32 | |
| 7 | 158.79 ; 0 | 0.0037 ; - | (N1) $\underline{N2}$ B (O) - $\underline{S1}$ (S2) | - | 72 | (FG) | 6 | 193.05 ; -1 | 0.0007 ; - | N1t1 N2t1 - (S1t$\underline{1}$) S2t1 | 0.034 | 95 | C1 |
| 8 | 157.45 ; - | 0.0051 ; - | N1 B - S1 (S$\underline{2}$) | - | 57 | (A') | 8 | 190.86 ; -1 | 0.0013 ; - | N1t2 N2t1 - (S1t$\underline{2}$) S2t2 | 0.030 | 91 | C2 |
| - | | | (N1) $\underline{N2}$ (B) - (S1) S2 | | 41 | | 10 | 186.03 ; - | 0.0045 ; - | P1t1 - S1t2 | - | 99 | B12 |
| | | | | | | | 12 | 181.59 ; 2 | 0.0363 ; -5 | P1t$\underline{2}$ - S1t1 | 0.137 | 96 | B21 |
| | | | | | | | 13 | 178.62 ; - | 0.0197 ; - | P1t1 - S2t1 | - | 53 | E1 |
| | | | | | | | - | | | (P1t2) P2t1 - S1t$\underline{1}$ (S2t2) | | 29 | |
| | | | | | | | 14 | 177.66 ; - | 0.0686 ; - | P2t1 - S1t1 | - | 46 | D1 |
| | | | | | | | + | | | P1t1 - (S1t2) S2t$\underline{1}$ | | | |
| | 3BSE dTpdT--dApdA (ref. to ideal dTpdT--dApdA) | | | | | | | 3BSE dApdTpdTpdG (ref. to ideal dApdTpdTpdG) | | | | | |
| 1 | 237.57 ; 1 | 0.0755 ; -81 | P1t1 - S1t1 | 0.069 | 74 | B1 | 1 | 240.60 ; 0 | 0.0494 ; -79 | P1t$\underline{2}$ - S1t2 | 0.065 | 81 | B2 |
| 2 | 233.73 ; 4 | 0.3469 ; 806 | P1t$\underline{2}$ - S1t2 | 0.080 | 70 | B2 | 3 | 234.53 ; 4 | 0.2908 ; 506 | P1t1 - S1t1 | 0.063 | 75 | B1 |
| 4 | 229.09 ; -2 | 0.0009 ; - | N1t1 - S1t1 | 0.015 | 83 | A1 | 4 | 232.87 ; -1 | 0.0010 ; - | N1t2 O - S1t2 | 0.054 | 96 | (A2) |
| 5 | 226.27 ; -1 | 0.0032 ; - | N1t2 - S1t2 (S2t2) | 0.031 | 95 | (A2) | 5 | 226.80 ; 2 | 0.0001 ; - | N1t1 (O) - S1t1 (S2t1) | 0.053 | 97 | (A1) |
| 13 | 193.23 ; 0 | 0.0006 ; - | (N1t1) N2t1 - (S1t$\underline{1}$) S2t1 | 0.047 | 96 | (C1) | 19 | 194.66 ; -4 | 0.0002 ; - | (N1t2) N2t2 (B) (O) - (S1t$\underline{2}$) S2t2 | 0.068 | 86 | (C2) |
| 15 | 191.59 ; -1 | 0.0028 ; - | N1t2 N2t2 - (S1t$\underline{2}$) S2t2 | 0.034 | 88 | (C2) | 23 | 190.65 ; 0 | 0.0020 ; - | (N1t1) N2t1 (B) (O) - (S1t$\underline{1}$) S2t1 | 0.101 | 82 | (C1) |
| 17 | 188.03 ; - | 0.0089 ; - | P2t1 A1 (A2) - S1t1 | 0.061 | 75 | (D1) | 27 | 188.88 ; - | 0.0167 ; - | P1t1 - S1t2 | - | 84 | B12 |
| 19 | 186.87 ; 2 | 0.0052 ; -51 | P2t1 (A1) A2 - S1t1 (A1) | - | 51 | (D1') | 32 | 183.95 ; - | 0.0282 ; - | P1t1 - S2t1 G | - | 83 | (E1) |
| - | | | A1 - (S1t1) A1 | | 27 | | 37 | 182.17 ; - | 0.0432 ; - | P1t2 (B) - S1t1 | - | 43 | (B21) |
| 20 | 185.89 ; - | 0.0193 ; - | (P1t2) (P2t1) (A1) A2 - S1t1 | - | 52 | (D1B 21) | + | | | B - G | | 37 | |
| + | | | P2t(1)2 (A2) - S1t(1)2 (A1) | | 27 | | 49 | 176.83 ; - | 0.0922 ; - | P2t2 (B) - S1t2 | - | 59 | (D2) |
| 22 | 184.98 ; 2 | 0.0261 ; 25 | P1t2 (B) - S1t1 | 0.082 | 83 | (B21) | + | | | B - S1t1 | | 16 | |

Following these rules, the transition expression of the excitations of each system can be determined. As an example, Figure 3 shows NTO1-H(E) of two excitations of di-Thy and dTpdT and the corresponding transition expressions. In addition, a similar categorization (Type-A~Type-G) introduced in the study of Thy and dT[95] for SO-hosted electronic excitations is also adopted here to classify electronic excitations: N1–S1 (Type-A); P1–S1 (Type-B); N1 N2–(S1) S2 (Type-C); P2–S1 (Type-D); P1–S2 (Type-E); N2–S1 (Type-F); N1 (N2)–S2 (Type-G). Since most of the studied segments here involve in two Thys, a SO-hosted excitation can be local (on the same Thy) or CT (across the two Thys) excitation. An excitation denoted as Type-A2 is therefore used to represent local Type-A excitation happened on the 2nd Thy. Type-B12, on the other hand, stands for a Type-B excitation with P1 orbital on the 1st Thy excited to the S1 orbital on the 2nd Thy. For simplicity, in the following discussions the terms "local Type-A(s)" will be used to refer to "local Type-A excitation(s)" and so on. Furthermore, an excitation can also be combined with different types of transitions. For instance, the excitation with NTO1 transition expression "P1t2 P2t1 - S1t1" is combined by Type-D1 local transition and Type-B21 CT transition with the opposite phase between P1t2 and P2t1 and is denoted as Type-D1B21.

We also employ $\sigma_E$ (environmentally resulted root mean square deviation of standard-orbitals coefficients) for examining the environmental effect to NTO1 transition origins of excitations of a core molecular segment, while for comparing results of the same molecules extracted from ideal and 3BSE sources it is denoted as $\sigma_G$ (geometry resulted deviation). The reference systems used for calculating different $\sigma$ are detailed in Tables 1 and 2.

### 3.2. Ideal Thy vs. di-Thy

To detail the environmentally resulted variation of excitation properties of several chosen core segments consisting of Thy(s), we take a deeper look into each type of electronic excitations. We first compare the results of ideal Thy and di-Thy shown in Table 1.

#### Local Type-B

Obviously the single bright local Type-B of Thy at 228.82nm evolves into two bright excitations of di-Thy: one being a little weakened at 232.87nm (Type-B2) and the other a little enhanced at 228.92nm (Type-B1). Moreover, the two bright excitations of di-Thy are both dominated by single NTO1 instead of by two similar weighted local transitions given in molecular exciton theory[98]. Therefore the separation of the two Thys is short such that their interaction cannot be well described by dipole-dipole interaction.

#### Local Type-A/C

As for the local Type-A, its absorption wavelength of Thy at 242.13nm results in the Type-A1 and Type-A2 of di-Thy at 244.16nm and 242.42nm; in other words, almost no energy splitting occurs. This is attributed to the weak transition dipole of Thy, which also leads to the small oscillator strengths. Similarly, no energy splitting occurs for the two weak local Type-C absorptions as well.

#### Local Type-D/E; CT Type-B

For even higher-lying excitations, Table 1(b) shows that they are mainly formed by local Type-D/E and CT Type-B transitions. The local Type-D/E involved excitations, like local Type-Bs mentioned above, have one brighter and the other darker and are both dominated by single NTO1. Moreover, their oscillator strengths are weakened by the mixture of CT Type-B transition. We can see that, for example, the brighter excitation involved with local Type-D transition of di-Thy at 179.55nm (Type-D1B21, f=0.0568) is still weaker than the pure Type-D of Thy at 178.49nm (f=0.0659).

The $\sigma_E$ for each local excitation shows that the Type-C1 of di-Thy has the largest deviation to be 0.113, but its transition expression feature remains intact.

### 3.3. Ideal di-Thy vs. dTpdT

We now move onto the effect of link between two Thys, i.e. the DNA backbone, on electronic excitations.

#### Local Type-B

In the presence of DNA backbone, the two bright absorption peaks of di-Thy become a brighter one and a darker one of dTpdT. The stronger one is predicted to be a Type-B1 (and with NTO2 "P1t2 - S1t2") at 236.23nm while the weaker one is a Type-B2 (with NTO2 "P1t1 - S1t1") at 239.30nm. Moreover, they both have a 3% energy lowering. Therefore it clearly shows that the presence of backbone is significant to modify the NTO1 transition origin of two original local Type-Bs of di-Thy, even if it does not donate orbitals to NTO1(2)-H(E). We find that the backbone here promotes the formation of exciton-like excitations only present for larger distance of two Thys, that is, $\frac{1}{\sqrt{2}}(\varphi_u^*\varphi_v \pm \varphi_u\varphi_v^*)$, where $\varphi_u^*$ and $\varphi_u$ is the excited and unexcited wavefunctions of monomer u and $\varphi_v^*$ and $\varphi_v$ those of monomer v of a dimer system composed of two geometrically identical monomers.

#### Local Type-A/C

Next we deal with the local Type-A. In the presence of backbone the two local Type-As instead have 3% and 4% energy arising, in contrast to the energy-lowering imposed on the local Type-Bs discussed above. The energy arising and lowering result in excitation reordering such that the lowest singlet excitation (Singlet-1) changes from a local Type-A to a local Type-B. Therefore, although still not taking into account more factors yet, e.g. base stacking, WC pairing nucleobases, ions, solvent molecules etc., the addition of backbone to di-Thy readily changes the excitation of Singlet-1. The presence of backbone also results in a slight S2-component mixing into the Type-A1 and Type-A2.

The remaining two $^1$nπ*-character excitations of di-Thy, i.e. local Type-Cs, though switching their relative order, retain their main feature of absorption energy, oscillator strength and transition origin of NTO1 in the presence of backbone.

#### Local Type-D/E; CT Type-B

Finally we consider the excitations formed by the mixture of CT Type-B transition and local Type-D/E transition. The Type-E1B12 has a strong enhancement of 72% of oscillator strength (from 0.0170 to 0.0292) with a small shifted excitation energy (188.49nm to 189.63nm), while the excitations of di-Thy with local Type-D transition (181.05nm, 179.55nm and 177.84nm) disappear in all the listed SO-excitations of dTpdT. Instead, they are replaced by excitations with local Type-E and/or CT Type-B transitions; that is, 186.74nm, 185.94nm and 179.79nm absorptions of dTpdT. Furthermore, the backbone orbital component is strongly involved in the 186.74nm and 185.94nm excitations. These results suggest that the larger average distance of electron transition which encompasses wider spatial area for these higher-lying excitations results in the high backbone-sensitivity.

### 3.4. Ideal dTpdT vs. dTpdT--dApdA
#### Local Type-B
For even larger system with dApdA in the opposing strand in charge of WC hydrogen bonding effect, the TD-ωB97X calculated absorption spectrum (Figure S3) shows that there are two strong peaks. However, after examining their NTO1 transition origins only the lower energy one at 239.09nm corresponds to a SO-hosted (Type-B1) excitation. Its NTO1 transition origin is very similar to that of the brightest peak of dTpdT and it has a minor energy shift (-1%). A similar situation also happens to the weaker Type-B2, the Singlet-1.

#### Local Type-A/C
The NTO1 transition origins of the two local Type-As are retained as well, but again their excitation energies arise (5% and 6%). For the two local Type-Cs, their absorption energies have minor deviation accompanied with a modest $\sigma_E$ (0.024 and 0.031).

#### Local Type-D/E; CT Type-B
For the CT Type-B involved excitations, interestingly the mixed local Type-E transition appearing in the result of dTpdT diminishes as dApdA is in effect, and two local Type-E involved excitations of dTpdT (186.74nm and 179.79nm) are replaced by local Type-D involved ones of dTpdT--dApdA (181.87nm and 180.96nm). The 180.96nm excitation even mixes in A2 in its transition expression. The remaining Type-E1B12 also has a quite large increase (101%) in its oscillator strength even if its absorption energy is rather intact. Therefore again environment effect is more pronounced in these higher-lying excitations.

### 3.5. Ideal dTpdT vs. dApdTpdTpdG
#### Local Type-B
In the presence of stacking nucleobases next to dTpdT and the backbone linking them, i.e. dApdTpdTpdG, the two local Type-Bs switch their positions such that the brighter absorption changes from the Type-B1 to the Type-B2, even if their excitation energies are not modified much. The oscillator strength of the brighter one is also dampened one half, while that of the darker one is enhanced. Hence it can be concluded that the local Type-Bs are quite sensitive to environmental surroundings of backbone and stacking nucleobases. They can be more exciton-like excitations, or can be well separated like two local ones. This can be comprehended that the environments experienced by the two Thys can mediate the degree of near-degeneracy of their local transitions so that the formation of exciton-like excitations can be tuned.

#### Local Type-A/C
On the other hand, the two local Type-As again have a 2% energy arising and a minor change of transition origin compared to the results of dTpdT.

As for the two local Type-Cs, the Type-C2 seems to be unspoiled in its transition origin but Type-C1 is strongly contaminated with backbone-orbital, also reflected in its large $\sigma_E$ (0.163); its NTO1 domination also decreases substantially. However, the energies of the two excitations are both rather intact.

#### Local Type-D/E; CT Type-B
The remaining excitations related to local Type-E and/or CT Type-B transitions of dTpdT again are affected quite intensively. The Type-E1B12, although retains similar excitation energy, have almost doubled its oscillator strength and its NTO1 transition expression mixes in A with S1t2 being diminished at the same time. On the other hand, the Type-E2B21 has more notable energy lowering (4%) but oscillator strength reduction (24%). The remaining Type-B21 and Type-E2B21 of dTpdT are replaced by the Type-F1 and Type-E2, both of which have noticeable other components, e.g. B, A or G, mixed in the transition expressions.

Therefore, the bases-stacking influence appears more pronounced than that of WC-pairing nucleobases through hydrogen bonding, as it leads to averagely larger $\sigma_E$.

### 3.6. Ideal vs. 3BSE (X-ray crystal) structures
We examine the effect of conformation variation on electronic excitation properties of dT, dTpdT, dTpdT--dApdA and dApdTpdTpdG. The transition origins of excitations of geometrically different systems can be compared with each other, given that the segment stays in the chemically same species.

#### dT
The detailed information of the first 10 (8 for dT) SO-hosted excitation of each 3BSE system is listed in Table 2. We can observe that for dT all 8 SO-hosted excitations except for the highest-lying Type-G can be well traced. A one-to-one mapping can then be readily made from their NTO1 transition origin. Moreover, their absorption energies, oscillator strengths and NTO1 transition origins are largely retained.

#### dTpdT
Analysis on dTpdT shows that the conformation variation is quite influential to the NTO1 transition origin of the two local Type-Bs such that the $\sigma_G$ value for Type-B1 is 0.221 and for Type-B2 is 0.126. The stronger absorption also transforms from the Type-B1 to the Type-B2, even if energy and oscillator strength of the stronger and weaker are modified little.

In contrast, the local Type-As and Type-Cs are rather sustainable in the NTO1 transition origin and absorption energy to conformation change, and their oscillator strengths remain small. Moreover, the Singlet-1 becomes Type-A1.

The remaining listed higher-lying excitations have rather large variations in their NTO1 transition origin although they are still around the absorption wavelength of 170nm~190nm; this can also be observed in the results of ideal di-Thy, dTpdT, dTpdT--dApdA and dApdTpdTpdG, that is, CT Type-B involved excitations are all located in this energy range, though their composition of NTO1 transition origin is very sensitive to the surrounding environment.

#### dTpdT--dApdA and dApdTpdTpdG
For much larger systems dTpdT--dApdA and dApdTpdTpdG, one may expect that conformation change accumulation should lead to quite distinct excitation properties. This seems to be the case for dApdTpdTpdG as only 6 excitations out of 10 can better correspond to excitations found in the result of ideal structure. Moreover, mixture of orbitals other than the standard-orbitals to transition origin often occurs. For example, Type-C1 and -C2 both have B and O involved in the transition expressions such that the ratio of local Type-C Thy-orbitals contributing to the total electronic density reduces. On the other hand, for dTpdT--dApdA there remains many excitations that can be well traced for the local Type-A, B and C transitions; the brighter and weaker local Type-Bs again exchange their NTO1 transition origin which should be due to their sensitivity to the coupling of the two Thys[97].

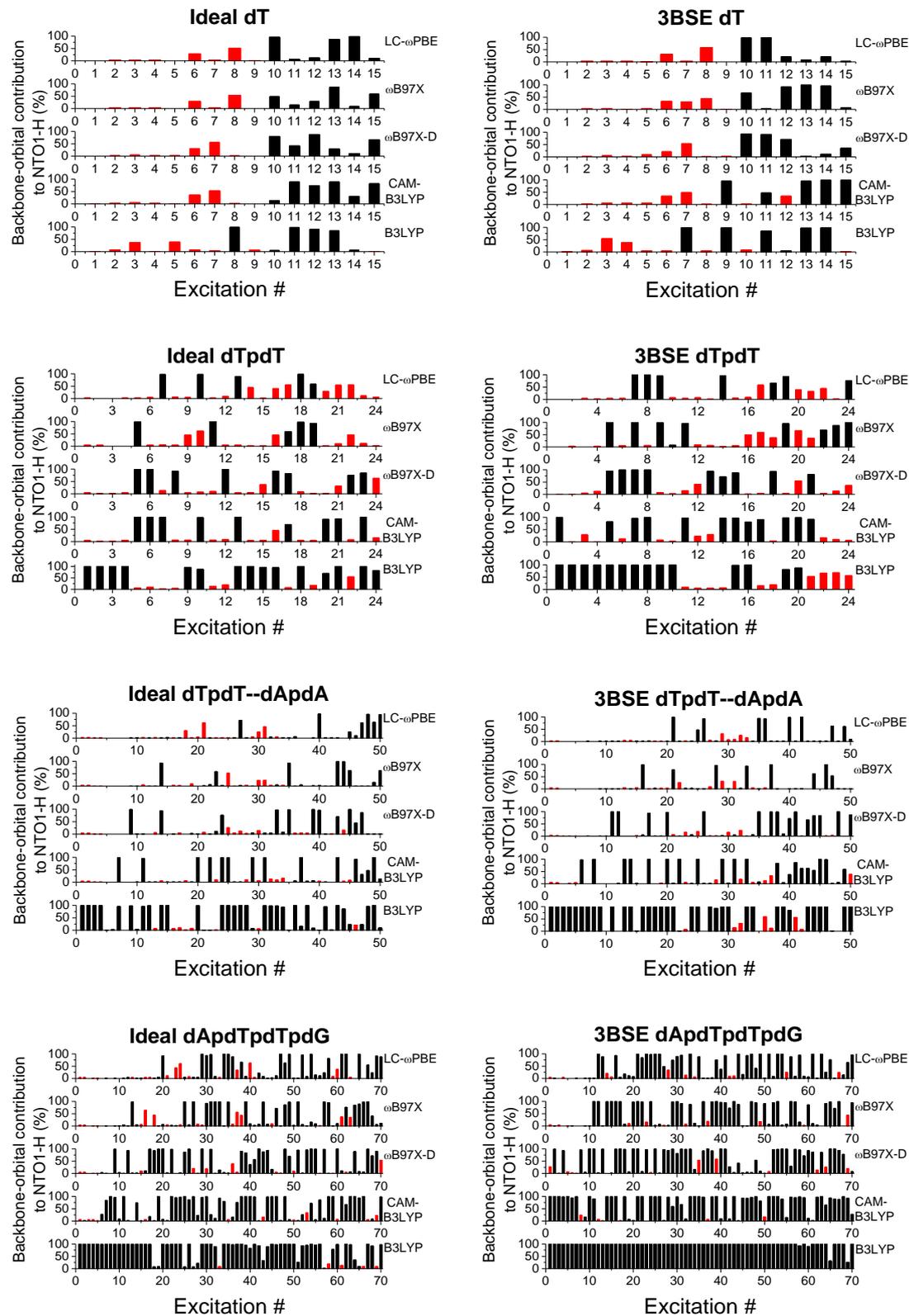

**Fig. 4** Percentage of backbone density contribution to the NTO1-H of the first several excitations of the backbone-included systems. Red columns indicate SO-hosted excitations.

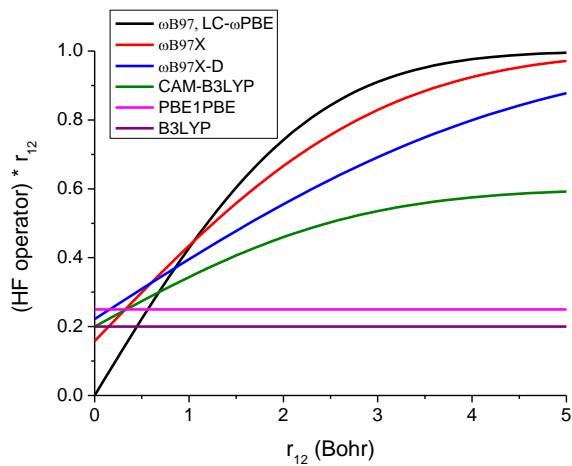

**Fig. 5** $L(r_{12})$ vs. $r_{12}$ for different functionals. For details, please see the text and Eqs. (1) ~ (3).

Overall, it can be seen that conformation influence can be the same or even outstrip environmental effects to the excitation properties of the SO-hosted excitations.

## 3.7. Detailed Backbone-Orbital Involvement in the First Several Excitations of B-DNA Segments Containing Backbone

Figure 4 shows the percentage of backbone-orbital density contributing to NTO1-H in the first 15, 24, 50, and 70 excitations (not only SO-hosted ones) of dT, dTpdT, dTpdT--dApdA, and dApdTpdTpdG, respectively. TD-B3LYP underestimates the absorption energy of a lot of backbone-to-base excitations crowding around the first several excitations. However, even LC hybrid functionals predict an impressive involvement of backbone-orbital in several excitations.

As shown in Eq. (2), the $L(r_{12})$ in a LC hybrid functional can be interpreted as a $r_{12}$-dependent proportion of HF exchange. For example, both ωB97 and LC-ωPBE adopt $L(r_{12}) = erf(\omega r_{12})$, with ω = 0.4 Bohr$^{-1}$, while ωB97X adopts $L(r_{12}) = a_x \times [1 - erf(\omega r_{12})] + erf(\omega r_{12})$, with ω = 0.3 Bohr$^{-1}$ and $a_x$ = 0.157706. By contrast, a global hybrid functional, containing a fixed proportion $c_x$ of HF exchange, adopts $L(r_{12}) = a_x$ and a pure DFA, containing no HF exchange, adopts $L(r_{12}) = 0$.

To look into more details, we plot, in Figure 5, the $r_{12}$-dependent proportion of Hartree-Fock exchange $L(r_{12})$, for various functionals. Evidently we observe that there is a strong correlation between the prescribed HF-exchange fraction in a LC hybrid functional and the appearance of significant backbone-to-base CT contributions to excitations. As the $L(r_{12})$ is larger for a functional in the range of inter-electronic separation $r_{12} > 1$ Bohr for the electron-hole interaction, the first excitation that has a significant contribution from backbone to NTO1-H appears earlier.

The growing up diversities among the results given by different LC hybrid functionals calculations (CAM-B3LYP[82], ωB97X[84], ωB97X-D[83] and LC-ωPBE[81]) due to the increase in the molecular size (dTpdT, dTpdT--dApdA, dApdTpdTpdG) suggest that the description of excitation order and transition origins is quite sensitive to the prescription of $L(r_{12})$ for larger and more complex systems. Therefore, finer tunings and revisions to the currently used functionals are necessary in order to achieve a more consistent description of these larger electronic systems for which more benchmarks are in need. Nevertheless, up to this point, we could still conclude that the DNA backbone is quantum mechanically quite important that its orbitals contribution is often strongly involved in the h-orbital of the electronic excitations of B-DNA.

## 4. Conclusion

We have examined several low-lying electronic excitations of various Thy-contained systems extracted from two sources of B-DNA, i.e. the ideal and a X-ray crystal structure. Various environmental and conformational effects on Thy-related excitations have been investigated quantum mechanically. The later and former effects are studied by comparing the LC-TDDFT results of two systems that are different from each other in geometry and in a physical/chemical surrounding, respectively. Through the employment of QNTO analysis, the transition origin of up to hundred electronic excitations of different systems can readily be determined and therefore similar excitations can be identified and compared with each other.

We find that although many low-lying SO-hosted $^1\pi\pi^*$- and $^1n\pi^*$-character excitations (or combinations of them) of Thy-contained systems can be interpreted by single NTO1, there are still several cases that NTO1 has less than 70% domination, in which case NTO2 is also taken into account. Therefore the present study provides a more realistic description of molecular exciton view in which the contribution of the local electronic transitions proportion can be different. Moreover, decomposition of NTO1(2)-H(E) of an excitation into standard-orbitals helps pinpoint its transition origin. It is possible that an excitation is a mixture of different local and CT transitions. It is also found that as the segment size grows to include in backbone or WC-pairing or adjacent stacking nucleobases, the NTO1 transition origin is often more complicated because various orbital-components are mixed; the NTO1 domination is also likely to decrease at the same time. Our analysis provides a quantitative investigation foundation for these complex phenomena.

Furthermore, both conformational and environmental effects to SO-hosted excitations are found to mediate not only the absorption energy and oscillator strength but also the NTO1 transition origin. This situation is significant especially for the higher-lying excitations composed of CT transitions. It is found that the higher-lying local Type-D, E and CT Type-B transitions are more sensitive to environmental surroundings as well as conformation variations than local Type-A, B and C transitions such that they often bear various mixtures to form the transition origin of excitations around 170nm~190nm. This situation probably results from their near-degenerate feature of transitions.

Overall, the present study suggests that a DNA molecule continually under various interactions with itself or environmental molecules in solution can exhibit from time to time a slightly or very different transition origin of NTO1(2) depending on its experience at the instance of light-absorptions. It has also been reviewed in the Chap. 14 of Ref. 36 that various electronic properties of nucleobases are notably dependent on their microscopic environment and present QNTO analysis helps further provide a systematic quantum chemistry framework to investigate them. Moreover, the origins of two electronic excitations of the same or different systems can be, as expressed by standard-orbitals, quite different even if their absorption energy and oscillator strength are similar to each other, which implies that a further determination of correct transition origin of an excitation is quite important which governs later dynamics of an electronically excited molecule.

Other possible effects such as solvents which are not considered in the present work are anticipated to be applicable in QNTO framework as well. The present scheme can also be readily extended to construct the potential energy surface of a gradually geometrically modified system within a complex environment because an electronic excited state can be well traced from its transition origin. More applications are underway.

## Acknowledgements

We would like to thank National Science Council (NSC), and National Center of Theoretical Sciences (NCTS) for their financial supports. We would also like to thank National Center for High-performance Computing (NCHC) for providing computational resources.

# Supporting Information

Significant Role of DNA Backbone in Mediating the Transition Origin of Electronic Excitations of B-DNA – Implication from Long Range Corrected TDDFT and Quantified NTO Analysis

Jian-Hao Li[a, b], Jeng-Da Chai[a, c, *], Guang-Yu Guo[a, c, d], Michitoshi Hayashi[b, *]

[a]Department of Physics, Center for Theoretical Sciences, National Taiwan University, Taipei, 10617, Taiwan
[b]Center for Condensed Matter Sciences, National Taiwan University, Taipei, 10617, Taiwan
[c]Center for Quantum Science and Engineering, National Taiwan University, Taipei, 10617, Taiwan
[d]Graduate Institute of Applied Physics, National Chengchi University, Taipei 11605, Taiwan

E-mail: jdchai@phys.ntu.edu.tw (J.-D. Chai); atmyh@ntu.edu.tw (M. Hayashi)

# Contents

**TABLE S1:** The major internal coordinates of DNA backbone – nucleic-acid torsion angles – of ideal and 3BSE dTpdT shown in Figure 1 and S1, adopted from the conventions defined in Chap. 5 of Ref. S1. The major differences of backbone bending of the two conformations evidently come from the deviations of $\delta_{T1}$, $\varepsilon_{T1}$ and $\beta_{T2}$. However, all the torsion-angles of the two dTpdTs are fairly within the common ranges of B-DNA structures, as can be referenced in p.132 of Ref. S1.

|  |  | Ideal dTpdT | 3BSE dTpdT |  |
|---|---|---|---|---|
| Angle | Sequence | Value(°) | Value(°) | Difference(°) |
| $\beta_{T1}$ | H*-O5'-C5'-C4' | -146.02040 | -171.79766 | -25.77726 |
| $\gamma_{T1}$ | O5'-C5'-C4'- | 36.36158 | 40.98660 | 4.62502 |
| $\delta_{T1}$ | C5'-C4'-C3'- | 156.40092 | 132.93283 | -23.46809 |
| $\varepsilon_{T1}$ | C4'-C3'-O3'-P | 154.98268 | -172.21242 | 32.8049 |
| $\zeta_{T1}$ | C3'-O3'-P-O5' | -95.17465 | -109.65894 | -14.48429 |
| $\alpha_{T2}$ | O3'-P-O5'-C5' | -46.83902 | -43.83471 | 3.00431 |
| $\beta_{T2}$ | P-O5'-C5'-C4' | -146.00015 | 172.24216 | -41.75769 |
| $\gamma_{T2}$ | O5'-C5'-C4'- | 36.37464 | 36.97245 | 0.59781 |
| $\delta_{T2}$ | C5'-C4'-C3'- | 156.39855 | 146.88308 | -9.51547 |
| $\varepsilon_{T2}$ | C4'-C3'-O3'-H* | 154.94931 | -115.74378 | 89.30691 |
| $\chi_{T1}$ | O4'-C1'-N1-C2 | -97.99829 | -110.05868 | -12.06039 |
| $\chi_{T2}$ | O4'-C1'-N1-C2 | -98.01020 | -97.36020 | 0.65000 |

H* denotes that P atom is replaced with H atom.

**TABLE S2:** The excitation properties of ideal di-Thy, dTpdT, dTpdT--dApdA, and dApdTpdG (Figure 1) provided by QNTO analysis and TDDFT with global hybrid (B3LYP[S2, S3] and PBE1PBE[S4]) and LC hybrid (CAM-B3LYP[S5], ωB97X-D[S6], and LC-ωPBE[S7]) functionals other than ωB97X[S8].  N denotes excitation order, whereas P denotes NTO2 phase for cases where NTO1 has less than 70% domination to the whole excitation.  λ (nm) and f stand for absorption wavelength and oscillator strength, respectively.  NTO1(2) and % record the expression of transition origin of the first (second) NTO pair and its domination to the whole excitation.  Type denotes the excitation classification based on the EOM-CCSD Thy NTO1 expressions used in Ref. S9.  The local Type-A, B and C derived excitations are marked in bold-face fond in the N column.  The data for ideal di-Thy and dTpdT is shown in (a), while that for ideal dTpdT--dApdA and dApdTpdTpdG is shown in (b).

| (a) | | Ideal di-Thy | | | | | Ideal dTpdT | | | | |
|---|---|---|---|---|---|---|---|---|---|---|---|
| | N | λ (nm) | f | NTO1 | % | Type | N; P | λ (nm) | f | NTO1(2) | % | Type |
| TD-B3LYP | **1** | 272.81 | 0.0001 | N1t1 - S1t1 | 98 | A1 | 5 | 274.25 | 0.0189 | P1t2 - S1t1 | 99 | B21 |
| | **2** | 272.37 | 0.0009 | N1t2 - S1t2 | 98 | A2 | 6 | 267.83 | 0.0316 | P1t1(2) - S1t2 | 98 | (B2B12) |
| | 3 | 267.89 | 0.0161 | P1t1 - S1t2 | 99 | B12 | **7** | 263.11 | 0.0005 | N1t1 - S1t1 | 99 | A1 |
| | 4 | 260.52 | 0.0232 | P1t2 - S1t1 | 99 | B21 | **8** | 262.12 | 0.0057 | N1t2 - S1t2 | 99 | A2 |
| | **5** | 249.98 | 0.0757 | P1t2 - S1t2 | 96 | B2 | **11** | 255.72 | 0.0203 | P1t12 (B) - S1t(1)2 | 83 | (B2) |
| | **6** | 243.91 | 0.1049 | P1t1(2) - S1t1 | 94 | (B1) | **12** | 253.85 | 0.2065 | P1t1(2) (B) - S1t1(2) | 80 | (B1) |
| | 7 | 238.23 | 0.0025 | N1t1 - S1t2 | 99 | A12 | 17 | 232.88 | 0.0008 | N1t2 - S1t1 | 100 | A21 |
| | 8 | 234.27 | 0.0017 | N1t2 - S1t1 | 99 | A21 | 19 | 228.83 | 0.0011 | N1t1 (B) - S1t2 | 95 | (A12) |
| | 9 | 220.52 | 0 | N2t2 - S1t2 | 95 | F2 | 22 | 224.93 | 0.0060 | N2t1 B - S1t1 | 93 | (F1) |
| | 10 | 215.67 | 0.0027 | P2t1(2) - S1t2 | 98 | (D2D12) | 25 | 211.56 | 0.0007 | (N1t1) (N2t1) B - S1t2 | 82 | (F12) |
| TD-PBE1PBE | **1** | 265.49 | 0 | N1t1 - S1t1 | 10 | A1 | 5 | 261.58 | 0.0309 | P1t(1)2 - S1t1 | 99 | (B1B21) |
| | **2** | 264.73 | 0.0002 | N1t2 - S1t2 | 99 | A2 | 6 | 257.30 | 0.0648 | P1t1(2) (N1t2) - S1t2 | 95 | (B2B12) |
| | 3 | 255.82 | 0.0349 | P1t(2) - S1t2 | 99 | (B2B12) | **7** | 255.94 | 0.0004 | N1t1 - S1t1 | 98 | A1 |
| | 4 | 248.99 | 0.0395 | P1t(1)2 - S1t1 | 99 | (B1B21) | **8** | 255.33 | 0.0255 | (P1t2) N1t2 - S1t2 | 97 | (A2) |
| | **5** | 243.88 | 0.0802 | P1t(1)2 - S1t2 | 97 | (B2) | **9** | 248.66 | 0.0354 | P1t(2) - S1t1(2) | 96 | (B1) |
| | **6** | 237.59 | 0.0932 | P1t(2) - S1t1 | 96 | (B1) | **10** | 247.18 | 0.1753 | P1t12 - S1t(1)2 | 93 | (B2) |
| | 7 | 223.87 | 0.0014 | N1t1 - S1t2 | 99 | A12 | 17 | 219.53 | 0.0006 | N1t2 - S1t1 | 97 | A21 |
| | 8 | 220.75 | 0.0012 | N1t2 - S1t1 | 99 | A21 | 18 | 217.66 | 0.0024 | N1t1 B - S1t2 | 80 | (A12) |
| | 9 | 211.45 | 0 | N2t2 - S1t2 | 92 | F2 | 19 | 216.99 | 0.0097 | (N1t1) (N2t2) B - S1t2 | 69 | (A12) |
| | 10 | 206.51 | 0.0113 | P2t1<u>2</u> - S1t2 | 95 | D2D12 | - | | | N2t1 B - S1t1 | 27 | |
| | | | | | | | 20 | 215.09 | 0.0021 | (N1t1) (N2t1) B - S1t12 | 72 | (AF1AF12) |
| TD-CAM-B3LYP | **1** | 251.57 | 0.0001 | N1t1 - S1t1 | 10 | A1 | **1** | 244.85 | 0.0289 | P1t(1)2 - S1t(1)2 | 66 | (B2) |
| | **2** | 249.81 | 0.0001 | N1t2 - S1t2 | 99 | A2 | - | | | P1t(2) - S1t1(2) | 31 | |
| | 3 | 238.34 | 0.1195 | P1t2 - S1t2 | 95 | B2 | **2** | 242.47 | 0.0027 | N1t2 - S1t2 | 97 | A2 |
| | 4 | 234.13 | 0.1854 | P1t1 - S1t1 | 95 | B1 | **3** | 242.10 | 0.0015 | N1t1 - S1t1 | 98 | A1 |
| | 5 | 210.53 | 0.0036 | P1t1 - S1t2 | 99 | B12 | **4** | 241.83 | 0.3783 | P1t1<u>2</u> - S1t1(<u>2</u>) | 64 | (B1) |
| | 6 | 201.23 | 0.0122 | P1t2 - S1t1 | 99 | B21 | + | | | P1t(1)2 - S1t(1)2 | 32 | |
| | 7 | 196.48 | 0.0000 | N1t2 N2t2 - (S1t<u>2</u>) S2t2 | 96 | C2 | 8 | 210.42 | 0.0052 | P1t2 - S1t1 | 75 | B21 |
| | 8 | 195.78 | 0.0010 | N1t1 N2t1 - (S1t<u>1</u>) S2t1 | 96 | C1 | 9 | 210.00 | 0.0206 | P1t1 - S1t2 | 76 | B12 |
| | 9 | 189.87 | 0.0304 | P2t2 - S1t2 | 94 | D2 | **11** | 197.09 | 0.0002 | N1t1 N2t1 - (S1t<u>1</u>) S2t1 | 94 | C1 |
| | 10 | 187.33 | 0.0664 | P2t1 - S1t1 | 90 | D1 | **12** | 193.18 | 0.0008 | N1t2 N2t2 - (S1t<u>2</u>) S2t2 | 92 | C2 |
| | | | | | | | 14 | 188.62 | 0.0689 | P1t2 - S2t2 | 82 | E2 |
| | | | | | | | 15 | 186.21 | 0.1807 | P1t1 - S2t1 | 69 | E1 |
| | | | | | | | - | | | P1t2 - (S1t<u>1</u>) S2t2 | 14 | |
| TD-ωB97X-D | **1** | 251.79 | 0.0001 | N1t1 - S1t1 | 99 | A1 | **1** | 245.02 | 0.0308 | P1t2 - S1t(1)2 | 67 | (B2) |
| | **2** | 250.28 | 0.0001 | N1t2 - S1t2 | 99 | A2 | - | | | P1t(2) - S1t1(2) | 30 | |
| | 3 | 238.31 | 0.1165 | P1t2 - S1t2 | 95 | B2 | **2** | 242.79 | 0.0014 | N1t2 - S1t2 | 98 | A2 |
| | 4 | 234.29 | 0.1851 | P1t1 - S1t1 | 95 | B1 | **3** | 242.20 | 0.0024 | N1t1 - S1t1 | 98 | A1 |
| | 5 | 207.47 | 0.0044 | P1t1 - S1t2 | 99 | B12 | **4** | 242.01 | 0.3728 | P1t(2) - S1t1(2) | 66 | (B1) |
| | 6 | 197.42 | 0.013 | P1t2 - S1t1 | 96 | B21 | + | | | P1t(1)2 - S1t(1)2 | 30 | |
| | 7 | 196.13 | 0 | N1t2 N2t2 - S1t2 S2t2 | 96 | (C2) | 7 | 207.83 | 0.0080 | P1t1 (B) - S1t2 | 98 | (B12) |
| | 8 | 195.44 | 0.0027 | N1t1 N2t1 - (S1t<u>1</u>) S2t1 | 92 | C1 | 9 | 205.77 | 0.0223 | P1t2 - S1t1 | 98 | B21 |
| | 9 | 190.73 | 0.0299 | P2t2 - S1t2 | 93 | D2 | **10** | 197.00 | 0.0002 | N1t1 N2t1 - (S1t<u>1</u>) S2t1 | 94 | C1 |
| | 10 | 188.28 | 0.0649 | P2t1 - S1t1 | 90 | D1 | **11** | 192.95 | 0.0006 | N1t2 N2t2 (B) - S1t<u>2</u> S2t2 | 91 | (C2) |
| | | | | | | | 13 | 188.64 | 0.0746 | P1t2 - S2t2 | 81 | E2 |
| | | | | | | | 14 | 186.38 | 0.1590 | P1t1 - S2t1 | 57 | E1 |
| | | | | | | | - | | | P2t1 - S1t1 | 27 | |
| TD-LC-ωPBE | **1** | 243.68 | 0.0001 | N1t1 - S1t1 (S2t1) | 99 | (A1) | **1** | 236.57 | 0.0361 | P1t2 - S1t2 | 66 | B2 |
| | **2** | 241.97 | 0.0001 | N1t2 - S1t2 (S2t2) | 99 | (A2) | - | | | P1t1 - S1t1 | 31 | |
| | 3 | 230.23 | 0.1279 | P1t2 - S1t2 | 94 | B2 | **2** | 235.8 | 0.0015 | N1t2 - S1t2 (S2t2) | 96 | (A2) |
| | 4 | 226.36 | 0.2347 | P1t1 - S1t1 | 93 | B1 | **3** | 234.58 | 0.0002 | N1t1 - S1t1 (S2t1) | 98 | (A1) |
| | 5 | 192.22 | 0 | (N1t2) N2t2 - (S1t<u>2</u>) S2t2 | 99 | (C2) | **4** | 233.55 | 0.4402 | P1t1 - S1t1 | 65 | B1 |
| | 6 | 191.55 | 0.0002 | N1t2 N2t1 - S2t1 | 99 | (C1) | + | | | P1t2 - S1t2 | 31 | |
| | 7 | 180.46 | 0.0349 | P1t1 - S1t2 S2t1 | 96 | E1B12 | 5 | 192.6 | 0 | (N1t1) N2t1 - S2t1 | 97 | (C1) |
| | 8 | 176.63 | 0.0256 | P2t2 - S1t2 | 85 | D2 | **6** | 188.95 | 0.0010 | (N1t2) N2t2 - (S1t<u>2</u>) S2t2 | 93 | (C2) |
| | 9 | 174.44 | 0.0887 | P2t1 - S1t1 | 84 | D1 | 8 | 182.85 | 0.0322 | P1t1 - S1t2 S2t1 | 83 | (E1B12) |
| | 10 | 172.45 | 0.1457 | P1t2 - (S1t1) S2t2 | 86 | (E2B21) | 9 | 180.62 | 0.2195 | P1t2 - (S1t1) S2t2 | 82 | (E2B21) |
| | | | | | | | 11 | 173.52 | 0.0856 | P1t2 (P2t<u>1</u>) - S1t1 (S2t<u>2</u>) | 87 | (D1E2B21) |
| | | | | | | | 12 | 172.38 | 0.0376 | P1t1 - S1t2 S2t<u>1</u> | 53 | (E1B12) |
| | | | | | | | - | | | P1t2 P2t1 - S1t1 | 40 | |

| (b) | | Ideal dTpdT--dApdA | | | | | | Ideal dApdTpdTpdG | | | | |
|---|---|---|---|---|---|---|---|---|---|---|---|---|
| | N;P | λ (nm) | f | NTO1(2) | % | Type | N;P | λ (nm) | f | NTO1(2) | % | Type |
| TD-CAM-B3LYP | 1 | 248.08 | 0.0246 | P1t(1)2 - S1t(1)2 | 69 | (B2) | 1 | 249.26 | 0.0606 | P1t1 - S1t1 | 86 | B1 |
| | - | | | P1t1(2) - S1t1(2) | 28 | | 3 | 244.73 | 0.2370 | P1t2 - S1t2 | 53 | B2 |
| | 2 | 245.12 | 0.3755 | P1t1(2) - S1t1 | 66 | (B1) | + | | | G - G | 36 | |
| | + | | | P1t(1)2 - S1t2 | 30 | | 4 | 237.64 | 0.0004 | N1t1 - S1t1 | 95 | A1 |
| | 3 | 231.39 | 0.0002 | N1t1 - S1t1 | 97 | A1 | 5 | 237.33 | 0.0002 | N1t2 - S1t2 | 97 | A2 |
| | 4 | 230.04 | 0.0001 | N1t2 - S1t2 | 97 | A2 | 16 | 221.42 | 0.0069 | P1t2 - S1t1 | 81 | B21 |
| | 14 | 214.76 | 0.0292 | P1t2 - S1t1 | 67 | B21 | 35 | 203.45 | 0.0117 | P1t1 - S1t2 | 90 | B12 |
| | + | | | A2 - S1t2 | 29 | | 43 | 197.10 | 0.0005 | N1t1 N2t1 (B) - S1t1 S2t1 | 82 | (C1) |
| | 18 | 206.77 | 0.0019 | P1t1 - S1t2 | 69 | B12 | 50 | 192.42 | 0.0400 | P1t2 - S2t2 | 68 | E2 |
| | + | | | A1 - A2 | 30 | | + | | | P1t1 (B) - (S1t2) G | 16 | |
| | 23 | 198.63 | 0 | (N1t1) N2t1 (B) - S1t1 S2t1 | 86 | (C1) | 53 | 191.51 | 0.0107 | (N1t2) (N2t2) B - (S1t2) S2t2 G | 49 | (C2) |
| | 28 | 195.30 | 0.0004 | (N1t2) N2t2 (B) - S1t2 S2t2 | 86 | (C2) | + | | | (N2t2) B - (S2t2) G | 38 | |
| | 30 | 191.51 | 0.0131 | P2t1 - S1t1 | 82 | D1 | 60 | 187.90 | 0.0864 | P1t12 - S2t1 | 59 | (E1E21) |
| | 32 | 190.77 | 0.0086 | N2t1 (P2t1) (B) (A1) - S1t1 (S2t1) | 76 | (F1D1) | + | | | P1t1(2) B - A | 28 | |
| TD-ωB97X-D | 1 | 248.30 | 0.0302 | P1t2 - S1t2 | 71 | B2 | 1 | 249.23 | 0.0587 | P1t1 - S1t1 | 85 | B1 |
| | 2 | 245.30 | 0.3651 | P1t1 - S1t1 | 69 | B1 | 2 | 245.46 | 0.1624 | P1t2 - S1t2 | 70 | B2 |
| | + | | | P1t2 - S1t2 | 26 | | 4 | 237.87 | 0.0002 | N1t1 - S1t1 | 96 | A1 |
| | 3 | 231.48 | 0.0001 | N1t1 - S1t1 | 97 | A1 | 5 | 237.56 | 0.0001 | N1t2 - S1t2 | 96 | A2 |
| | 4 | 230.35 | 0.0001 | N1t2 - S1t2 | 97 | A2 | 15 | 216.21 | 0.0296 | P1t2 (B) - S1t1 | 96 | B21 |
| | 13 | 210.21 | 0.0148 | P1t2 - S1t1 | 99 | B21 | 27 | 200.77 | 0.0099 | P1t1 (B) - S1t2 (A) | 93 | (B12) |
| | 17 | 204.45 | 0.0030 | P1t1 - S1t2 | 76 | B12 | 30 | 197.19 | 0.0002 | N1t1 N2t1 (B) - S1t1 S2t1 | 79 | (C1) |
| | 21 | 198.53 | 0.0002 | (N1t1) N2t1 A2 - S1t1 (S2t1) | 68 | (C1) | 36 | 192.25 | 0.0023 | (N1t2) N2t2 B - S1t2 S2t2 | 73 | (C2) |
| | + | | | (N1t1) (N2t1) A2 - S2t1 A1 | 21 | | 37 | 191.61 | 0.0732 | P1t2 - S2t2 | 75 | E2 |
| | 25 | 195.18 | 0.0007 | (N1t2) N2t2 (B) - S1t2 (S2t2) | 84 | (C2) | 46 | 186.99 | 0.1200 | P1t1(2) - (S1t2) S2t1 | 71 | (E1B12E21) |
| | 26 | 192.00 | 0.0152 | P2t1 - S1t1 | 82 | D1 | | | | | | |
| | 27 | 191.06 | 0.0093 | N2t1 (P2t1) (B) - S1t1 (S2t1) | 81 | (F1D1) | | | | | | |
| TD-LC-ωPBE | 1 | 239.11 | 0.0366 | P1t2 - S1t2 | 70 | B2 | 1 | 240.07 | 0.0436 | P1t1 - S1t1 | 75 | B1 |
| | 2 | 236.36 | 0.4199 | P1t1 - S1t1 | 69 | B1 | 2 | 236.91 | 0.2569 | P1t2 - S1t2 | 75 | B2 |
| | + | | | P1t2 - S1t2 | 26 | | 4 | 231.02 | 0.0001 | N1t2 - S1t2 (S2t2) | 96 | (A2) |
| | 3 | 223.01 | 0.0001 | N1t1 - S1t1 (S2t1) | 95 | (A1) | 5 | 230.58 | 0.0002 | N1t1 - S1t1 (S2t1) | 95 | (A1) |
| | 4 | 222.54 | 0.0001 | N1t2 - S1t2 (S2t2) | 96 | (A2) | 13 | 191.95 | 0 | (N1t1) N2t1 - (S1t1) S2t1 | 94 | (C1) |
| | 11 | 193.57 | 0 | (N1t1) N2t1 - S2t1 | 95 | (C1) | 16 | 187.27 | 0.0080 | (N1t2) N2t2 - S2t2 | 86 | (C2) |
| | 13 | 190.56 | 0.0003 | (N1t2) N2t2 - S2t2 | 95 | (C2) | 17 | 186.33 | 0.0405 | P1t2 - S1t1 S2t1 | 67 | E2B21 |
| | 16 | 182.39 | 0.0216 | (P1t2) P2t1 (A1) - S1t1 | 62 | (D1B21) | - | | | A - A | 11 | |
| | + | | | A1 (A2) - A1 | 12 | | 21 | 183.11 | 0.0967 | P1t1 (B) - (S1t2) S2t1 (A) | 71 | (E1B12) |
| | 18 | 181.19 | 0.0461 | P1t1 B - S1t2 (S2t1) | 71 | E1B12 | 23 | 180.92 | 0.0294 | P1t2 B - S1t1 (G) | 77 | (B21) |
| | 19 | 180.79 | 0.0006 | (P1t2) (P2t2) A1 (A2) - S1t1(2) (S2t2) (A2) | 41 | - | 24 | 180.36 | 0.0067 | P1t2 B - S1t1 | 82 | (B21) |
| | + | | | P2t2 (A1) (A2) - S1t(1)2 (A2) | 27 | | | | | | | |
| | 20 | 180.07 | 0.0412 | P1t2 - S1t1 (S2t2) | 43 | (E2B21) | | | | | | |
| | + | | | B - S1t2 | 26 | | | | | | | |

S4

**TABLE S3:** The NTO1(2) of the first 10 SO-hosted singlet excitations of ideal dTpdT calculated by TD-ωB97X with large basis set 6-311+G(df,p). The excitation of each system is referenced to that of the ideal system (referred to as the core molecular unit) shown in the subtitle recording the system name if an excitation correspondence is identified. For instance, the 3$^{rd}$ excitation (Type-A2) of dTpdT is referenced to the 2$^{nd}$ excitation of di-Thy, while the 9$^{th}$ excitation (Type-E2B21) of dTpdT, having no similar type of excitation in di-Thy, is not referenced to any excitation. The transition origin variation is denoted as $\sigma_B$ (basis set resulted deviation). The local Type-A, B and C derived excitations that can be clearly recognized are marked with shadow background in the N column.

| N; P | λ (nm) ; Diff. (%) of Energy | | f ; Diff. (%) of f | | NTO1 | $\sigma_B$ | % | Type |
|---|---|---|---|---|---|---|---|---|
| 1 | 248.72 | -4 | 0.0331 | -11 | P1t2 - S1t2 | 0.019 | 62 | B2 |
| - | | | | | P1t1 - S1t1 | | 34 | |
| 2 | 245.49 | -4 | 0.4171 | 0 | P1t1(2) - S1t1(2) | 0.044 | 64 | (B1) |
| + | | | | | P1t(1)2 - S1t(1)2 | | 32 | |
| 3 | 236.75 | 0 | 0.0001 | - | N1t2 - S1t2 (S2t2) | 0.021 | 95 | (A2) |
| 4 | 236.56 | -1 | 0 | - | N1t1 - S1t1 | 0.010 | 96 | A1 |
| 6 | 194.33 | -2 | 0.0169 | -42 | P1t1 (B) - S1t2 S2t1 | 0.033 | 81 | (E1B12) |
| 7 | 193.20 | -1 | 0.0013 | - | N1t1 N2t1 - (S1t1) S2t1 | 0.027 | 91 | C1 |
| 8 | 192.30 | -3 | 0.1237 | 159 | P1t2 - S1t1 S2t2 | 0.119 | 85 | E2B21 |
| 10 | 189.92 | -1 | 0.0050 | - | (N1t2) N2t2 - (S1t2) S2t2 | 0.057 | 64 | (C2) |
| + | | | | | B - O | | 21 | |
| 13 | 184.59 | -3 | 0.0782 | -14 | P1t2 - S1t1 S2t2 (O) | 0.036 | 88 | (E2B21) |
| 15 | 182.63 | - | 0.0978 | - | P1t1 - S1t2 S2t1 (O) | - | 56 | (E1B12) |
| - | | | | | P2t1 B - S1t1 | | 36 | |

**Basis Set Dependence of TD-ωB97X Calculation for dTpdT – 6-31G(d) vs. 6-311+G(df,p)**

As shown in Table S3, the first 9 SO-hosted excitations given by 6-311+G(df,p) calculation each has a corresponding one in the 6-31G(d) result (Table 1(b)) and only the Type-E2B21 among them has a rather pronounced $\sigma_B$ to be 0.119, which is also reflected in its transition expression of 6-31G(d) result bearing additional backbone-orbital involvement. As for the excitation order, we see that difference mainly arises from the CT Type-B involved excitations, i.e. switch of Type-E1B12 and Type-C1, and Type-E2B21 and Type-C2; the higher-lying Type-B21 is also replaced by a Type-E1B12. With regard to excitation energy, the local Type-As and Type-Cs ($^1$nπ*-character) have values almost unchanged, while the two local Type-Bs ($^1$ππ*-character) both have 4% lowering and the remaining CT Type-B involved excitations have 2~3% lowering. For oscillation strength it is much retained for local Type-Bs, one of which is responsible for the brightest peak (Type-B1). On the other hand, the local Type-A, C and CT Type-B excitations have larger deviation of oscillator strength (but remain dark absorptions) except the Type-E2B21 which instead has a 159% enhancement, probably due to the 6-311+G(df,p) result excluding the backbone-orbital involvement in its transition origin.

Overall, the result of local Type-A, B and C excitations given by 6-31G(d) calculation is quite similar to those given by the much more expensive 6-311G+(df,p) calculation; more deviations come from results of CT Type-B involved excitations which have larger average electron transition distance.

**TABLE S4:** The detailed standard-orbitals projection coefficients or coefficients square of NTO1(2)-H(E) of several SO-hosted excitations of (a) ideal di-Thy, dTpdT, dTpdT--dApdA, and dApdTpdTpdG and (b) 3BSE dT, dTpdT, dTpdT--dApdA, and dApdTpdTpdG calculated by TD-ωB97X. For orbital-fraction from backbone, the coefficients square which corresponds to electronic density contribution is shown instead. (c) The data for the coefficients square of orbitals from adenine and/or guanine in the ideal dTpdT--dApdA and dApdTpdTpdG cases.

| (a) | λ (nm) | h-orbital | | | | | | | | | e-orbital | | | |
|---|---|---|---|---|---|---|---|---|---|---|---|---|---|---|
| | | P1$_{T1}$ | P1$_{T2}$ | N1$_{T1}$ | N1$_{T2}$ | P2$_{T1}$ | P2$_{T2}$ | N2$_{T1}$ | N2$_{T2}$ | (B)$^2$ | S1$_{T1}$ | S1$_{T2}$ | S2$_{T1}$ | S2$_{T2}$ |
| Ideal di-Thy | 244.16 | -0.04 | 0.00 | 0.96 | 0.00 | 0.00 | 0.01 | -0.29 | 0.00 | | 0.93 | 0.02 | 0.31 | 0.03 |
| | 242.42 | -0.01 | -0.04 | 0.00 | 0.96 | 0.00 | -0.01 | 0.01 | -0.27 | | 0.03 | 0.93 | 0.06 | 0.31 |
| | 232.87 | -0.07 | 0.99 | -0.01 | 0.04 | 0.00 | -0.05 | -0.02 | 0.00 | | 0.02 | 0.99 | 0.08 | -0.03 |
| | 228.92 | 0.99 | -0.10 | 0.04 | -0.01 | 0.00 | -0.01 | 0.00 | 0.01 | | 0.99 | -0.05 | -0.07 | 0.04 |
| | 191.44 | 0.01 | 0.00 | 0.01 | 0.57 | 0.01 | 0.00 | -0.02 | 0.82 | | 0.04 | -0.39 | -0.02 | 0.89 |
| | 190.81 | -0.10 | -0.02 | 0.60 | -0.01 | 0.00 | 0.00 | 0.78 | 0.01 | | -0.32 | 0.19 | 0.90 | 0.00 |
| | 188.49 | 0.99 | 0.05 | 0.05 | 0.01 | -0.09 | 0.06 | 0.05 | 0.01 | | 0.04 | 0.89 | 0.45 | 0.03 |
| | 181.05 | -0.05 | 0.07 | 0.02 | 0.01 | -0.06 | 0.99 | 0.02 | 0.00 | | 0.06 | 0.99 | 0.10 | 0.07 |
| | 179.55 | 0.05 | 0.71 | 0.00 | 0.02 | -0.69 | 0.11 | -0.03 | 0.01 | | 0.95 | -0.01 | 0.00 | 0.29 |
| | 177.84 | 0.08 | 0.69 | -0.01 | 0.03 | 0.71 | 0.07 | -0.03 | -0.02 | | 0.94 | 0.03 | -0.02 | 0.33 |
| Ideal dTpdT | 239.30 | -0.20 | 0.95 | 0.01 | 0.07 | 0.01 | -0.01 | -0.02 | -0.02 | 0.04 | 0.21 | 0.96 | 0.09 | 0.02 |
| | | 0.95 | 0.21 | -0.01 | 0.02 | -0.02 | 0.01 | 0.01 | -0.01 | 0.04 | 0.96 | -0.21 | -0.04 | 0.03 |
| | 236.23 | 0.94 | -0.24 | 0.00 | -0.03 | -0.02 | -0.01 | 0.01 | 0.02 | 0.04 | 0.96 | -0.23 | -0.05 | 0.06 |
| | | 0.26 | 0.92 | -0.02 | 0.17 | -0.04 | -0.04 | -0.01 | -0.04 | 0.04 | 0.23 | 0.95 | 0.08 | -0.03 |
| | 235.97 | -0.01 | -0.10 | 0.01 | 0.95 | 0.01 | 0.00 | 0.01 | -0.27 | 0.00 | 0.03 | 0.91 | 0.06 | 0.33 |
| | 235.06 | -0.01 | 0.00 | 0.96 | -0.01 | 0.00 | 0.01 | -0.27 | 0.00 | 0.00 | 0.92 | 0.02 | 0.32 | 0.04 |
| | 192.01 | -0.03 | -0.01 | 0.56 | -0.01 | 0.00 | 0.00 | 0.79 | 0.01 | 0.04 | -0.36 | 0.14 | 0.89 | 0.01 |
| | 189.63 | 0.96 | 0.02 | 0.02 | 0.00 | -0.08 | 0.02 | 0.01 | -0.03 | 0.05 | -0.01 | 0.82 | 0.56 | 0.07 |
| | 188.34 | -0.02 | 0.13 | 0.00 | 0.56 | -0.01 | 0.00 | -0.01 | 0.78 | 0.04 | 0.05 | -0.34 | -0.01 | 0.90 |
| | 186.74 | 0.05 | 0.72 | -0.01 | -0.01 | -0.05 | 0.09 | 0.01 | -0.04 | 0.45 | 0.92 | 0.02 | 0.00 | 0.36 |
| | 185.94 | 0.10 | 0.59 | 0.00 | 0.01 | -0.10 | 0.08 | -0.04 | -0.01 | 0.61 | 0.94 | 0.04 | 0.02 | 0.29 |
| | 179.79 | -0.02 | 0.97 | -0.01 | 0.00 | -0.08 | 0.06 | -0.01 | -0.01 | 0.04 | 0.59 | -0.01 | -0.08 | -0.78 |
| Ideal dTpdT (Large Basis) | 248.72 | -0.21 | 0.94 | -0.01 | 0.03 | 0.00 | 0.01 | -0.01 | 0.01 | 0.02 | 0.22 | 0.94 | 0.10 | 0.03 |
| | | 0.95 | 0.23 | 0.00 | -0.02 | 0.02 | 0.00 | 0.00 | 0.00 | 0.04 | 0.95 | -0.23 | -0.04 | 0.04 |
| | 245.49 | 0.90 | -0.33 | -0.01 | -0.03 | 0.02 | -0.01 | 0.01 | 0.00 | 0.03 | 0.92 | -0.33 | -0.05 | 0.07 |
| | | 0.35 | 0.91 | 0.00 | 0.04 | -0.04 | -0.02 | -0.01 | 0.00 | 0.04 | 0.33 | 0.91 | 0.10 | -0.03 |
| | 236.75 | -0.02 | -0.03 | 0.01 | 0.96 | 0.01 | 0.00 | 0.01 | -0.26 | 0.00 | 0.03 | 0.91 | 0.06 | 0.32 |
| | 236.56 | -0.01 | 0.00 | 0.96 | -0.02 | 0.02 | 0.00 | -0.25 | 0.00 | 0.00 | 0.91 | 0.01 | 0.31 | 0.04 |
| | 194.33 | 0.93 | 0.01 | -0.05 | 0.01 | -0.07 | 0.02 | -0.06 | -0.01 | 0.11 | -0.02 | 0.79 | 0.58 | 0.06 |
| | 193.20 | 0.05 | -0.03 | 0.56 | -0.02 | -0.01 | 0.00 | 0.79 | 0.00 | 0.06 | -0.38 | 0.16 | 0.86 | 0.00 |
| | 192.30 | 0.11 | 0.96 | 0.01 | 0.07 | -0.06 | 0.09 | 0.00 | 0.06 | 0.02 | 0.76 | 0.07 | 0.00 | 0.61 |
| | 189.92 | -0.01 | -0.04 | 0.00 | 0.54 | -0.01 | -0.01 | -0.01 | 0.78 | 0.08 | 0.06 | -0.42 | -0.02 | 0.85 |
| | | 0.16 | -0.03 | 0.00 | 0.07 | -0.01 | 0.00 | 0.00 | 0.12 | 0.95 | -0.15 | 0.19 | 0.03 | 0.19 |
| | 184.59 | -0.01 | 0.96 | 0.00 | 0.01 | -0.05 | 0.06 | 0.01 | -0.02 | 0.05 | 0.65 | -0.01 | -0.07 | -0.68 |
| | 182.63 | 0.96 | -0.06 | 0.01 | -0.02 | 0.03 | 0.02 | -0.03 | 0.00 | 0.04 | -0.05 | 0.58 | -0.74 | 0.03 |
| | | -0.02 | 0.14 | 0.00 | 0.01 | 0.70 | -0.01 | 0.00 | 0.01 | 0.47 | 0.96 | 0.05 | -0.01 | 0.08 |
| Ideal dTpdT--dApdA | 242.05 | -0.15 | 0.95 | -0.01 | 0.02 | 0.00 | -0.09 | -0.01 | 0.01 | 0.04 | 0.15 | 0.96 | 0.08 | 0.08 |
| | 239.09 | 0.95 | -0.17 | 0.01 | -0.01 | -0.09 | 0.00 | 0.01 | 0.00 | 0.04 | 0.98 | -0.13 | 0.01 | 0.04 |
| | 224.02 | 0.00 | 0.01 | 0.94 | 0.00 | 0.00 | 0.00 | -0.22 | 0.00 | 0.01 | 0.92 | 0.04 | 0.32 | 0.04 |
| | 223.27 | -0.02 | -0.02 | 0.00 | 0.94 | 0.02 | 0.00 | 0.00 | -0.22 | 0.01 | 0.04 | 0.91 | 0.09 | 0.33 |
| | 193.00 | 0.02 | -0.02 | 0.52 | -0.01 | -0.01 | 0.03 | 0.81 | 0.01 | 0.05 | -0.37 | 0.17 | 0.88 | 0.01 |
| | 189.76 | -0.03 | 0.04 | 0.01 | 0.52 | 0.01 | -0.01 | 0.00 | 0.81 | 0.04 | 0.06 | -0.36 | -0.01 | 0.90 |
| | 188.97 | 0.08 | 0.91 | 0.01 | -0.01 | -0.13 | 0.07 | -0.01 | -0.04 | 0.03 | 0.94 | 0.00 | 0.04 | 0.30 |
| | 186.79 | 0.92 | 0.03 | 0.00 | 0.01 | -0.19 | 0.03 | -0.01 | 0.00 | 0.09 | -0.01 | 0.84 | 0.51 | 0.08 |
| | 181.87 | 0.10 | 0.18 | -0.04 | 0.01 | 0.57 | -0.01 | -0.16 | -0.01 | 0.52 | 0.97 | 0.02 | 0.09 | 0.08 |
| | 180.96 | -0.05 | 0.08 | 0.02 | -0.01 | 0.00 | 0.71 | 0.04 | -0.01 | 0.01 | 0.17 | 0.93 | 0.13 | 0.19 |
| Ideal dApdTpdTpdG | 242.97 | 0.94 | -0.17 | 0.03 | -0.03 | -0.03 | -0.01 | 0.01 | 0.01 | 0.04 | 0.97 | 0.12 | 0.02 | 0.05 |
| | 239.62 | -0.12 | 0.96 | 0.00 | 0.04 | 0.00 | -0.01 | -0.01 | 0.00 | 0.04 | -0.12 | 0.97 | 0.10 | -0.03 |
| | 231.11 | -0.02 | -0.02 | 0.02 | 0.96 | 0.01 | 0.01 | 0.00 | -0.27 | 0.01 | 0.04 | 0.90 | 0.07 | 0.33 |
| | 230.97 | -0.02 | 0.00 | 0.96 | -0.03 | 0.01 | 0.00 | -0.26 | 0.01 | 0.01 | 0.91 | 0.03 | 0.33 | 0.04 |
| | 194.60 | 0.10 | 0.95 | 0.00 | 0.03 | -0.05 | 0.11 | -0.02 | 0.01 | 0.06 | 0.90 | 0.00 | 0.02 | 0.36 |
| | 192.01 | -0.04 | -0.10 | 0.31 | 0.00 | 0.03 | 0.00 | 0.49 | 0.00 | 0.64 | 0.88 | 0.00 | -0.42 | 0.04 |
| | | 0.00 | 0.07 | 0.48 | -0.02 | -0.03 | 0.00 | 0.61 | -0.01 | 0.39 | 0.38 | 0.18 | 0.85 | 0.06 |
| | 190.91 | 0.12 | 0.04 | 0.43 | -0.01 | -0.01 | 0.01 | 0.59 | -0.01 | 0.44 | -0.75 | 0.14 | 0.60 | -0.03 |
| | | 0.09 | -0.10 | 0.34 | -0.01 | 0.05 | -0.01 | 0.52 | 0.00 | 0.58 | 0.58 | 0.00 | 0.74 | 0.08 |
| | 188.45 | 0.95 | -0.08 | -0.01 | -0.03 | -0.02 | 0.01 | -0.03 | 0.00 | 0.05 | -0.03 | 0.47 | 0.75 | 0.06 |
| | 186.62 | 0.01 | 0.03 | 0.01 | 0.56 | -0.01 | -0.02 | 0.00 | 0.78 | 0.05 | 0.04 | -0.35 | -0.02 | 0.90 |
| | 182.31 | -0.05 | 0.95 | 0.00 | 0.00 | -0.02 | 0.01 | -0.01 | -0.02 | 0.08 | -0.26 | -0.01 | 0.03 | 0.80 |
| | | 0.76 | 0.14 | 0.00 | -0.04 | -0.08 | 0.15 | 0.01 | -0.03 | 0.32 | -0.08 | 0.58 | 0.24 | 0.25 |

| (b) | λ (nm) | h-orbital | | | | | | | | | e-orbital | | | |
|---|---|---|---|---|---|---|---|---|---|---|---|---|---|---|
| | | P1$_{T1}$ | P1$_{T2}$ | N1$_{T1}$ | N1$_{T2}$ | P2$_{T1}$ | P2$_{T2}$ | N2$_{T1}$ | N2$_{T2}$ | (B)$^2$ | S1$_{T1}$ | S1$_{T2}$ | S2$_{T1}$ | S2$_{T2}$ |
| 3BSE dT | 238.04 | 0.00 | 0.04 | 0.00 | 0.95 | 0.00 | -0.02 | 0.00 | -0.31 | 0.00 | 0.00 | 0.92 | 0.00 | 0.32 |
| | 228.90 | 0.00 | 0.98 | 0.00 | -0.04 | 0.00 | -0.02 | 0.00 | 0.02 | 0.03 | 0.00 | 0.99 | 0.00 | -0.04 |
| | 191.10 | 0.00 | 0.09 | 0.00 | 0.59 | 0.00 | 0.05 | 0.00 | 0.77 | 0.03 | 0.00 | -0.34 | 0.00 | 0.90 |
| | 176.81 | 0.00 | 0.07 | 0.00 | 0.02 | 0.00 | 0.98 | 0.00 | 0.03 | 0.01 | 0.00 | 0.99 | 0.00 | 0.11 |
| | 176.06 | 0.00 | 0.98 | 0.00 | -0.06 | 0.00 | 0.00 | 0.00 | -0.05 | 0.03 | 0.00 | 0.02 | 0.00 | 0.99 |
| | 168.14 | 0.00 | 0.00 | 0.00 | 0.22 | 0.00 | -0.06 | 0.00 | 0.73 | 0.33 | 0.00 | 0.95 | 0.00 | 0.25 |
| | 158.79 | 0.00 | 0.02 | 0.00 | 0.40 | 0.00 | -0.01 | 0.00 | -0.65 | 0.31 | 0.00 | -0.85 | 0.00 | 0.48 |
| | 157.45 | 0.00 | -0.01 | 0.00 | 0.72 | 0.00 | -0.03 | 0.00 | 0.04 | 0.44 | 0.00 | 0.86 | 0.00 | -0.44 |
| | | 0.00 | 0.08 | 0.00 | 0.37 | 0.00 | -0.10 | 0.00 | -0.85 | 0.11 | 0.00 | 0.46 | 0.00 | 0.86 |
| 3BSE dTpdT | 238.34 | -0.04 | 0.00 | 0.95 | 0.00 | 0.02 | 0.01 | -0.28 | 0.00 | 0.00 | 0.92 | 0.02 | 0.32 | 0.00 |
| | 235.85 | 0.84 | -0.43 | 0.07 | -0.21 | -0.03 | -0.02 | -0.01 | 0.07 | 0.03 | 0.87 | 0.47 | -0.02 | 0.05 |
| | | 0.48 | 0.73 | 0.04 | 0.41 | 0.00 | -0.03 | -0.01 | -0.13 | 0.03 | -0.48 | 0.86 | 0.05 | 0.08 |
| | 234.90 | 0.00 | -0.13 | 0.00 | 0.94 | 0.00 | -0.03 | 0.00 | -0.30 | 0.00 | 0.01 | 0.92 | 0.01 | 0.32 |
| | 231.25 | -0.19 | 0.95 | 0.00 | -0.10 | 0.02 | -0.03 | -0.01 | 0.02 | 0.03 | -0.18 | 0.97 | 0.03 | -0.05 |
| | | 0.95 | 0.19 | -0.03 | -0.04 | -0.05 | -0.02 | 0.01 | 0.00 | 0.04 | 0.97 | 0.18 | 0.01 | 0.00 |
| | 193.05 | -0.02 | 0.00 | 0.56 | 0.00 | -0.01 | 0.00 | 0.78 | -0.01 | 0.05 | -0.36 | 0.03 | 0.90 | -0.01 |
| | 190.86 | 0.00 | 0.10 | 0.00 | 0.57 | 0.01 | 0.05 | 0.00 | 0.76 | 0.06 | 0.01 | -0.40 | 0.00 | 0.88 |
| | 186.03 | 0.95 | 0.05 | -0.01 | 0.01 | -0.05 | -0.04 | -0.01 | 0.01 | 0.07 | 0.01 | 0.96 | 0.23 | 0.04 |

S6

|  |  |  |  |  |  |  |  |  |  |  |  |  |  |
|---|---|---|---|---|---|---|---|---|---|---|---|---|---|
|  | 181.59 | 0.07 | 0.92 | -0.01 | 0.02 | 0.23 | 0.11 | 0.00 | -0.01 | 0.08 | 0.94 | 0.01 | -0.02 | 0.30 |
|  | 178.62 | 0.91 | -0.03 | 0.00 | 0.01 | 0.26 | 0.01 | -0.06 | 0.02 | 0.06 | 0.26 | -0.23 | 0.92 | 0.04 |
|  |  | -0.27 | 0.47 | -0.05 | 0.00 | 0.78 | 0.05 | -0.11 | -0.01 | 0.06 | -0.89 | -0.08 | 0.22 | 0.37 |
|  | 177.66 | -0.29 | -0.22 | -0.01 | -0.01 | 0.92 | 0.03 | 0.03 | 0.00 | 0.01 | 0.93 | -0.09 | 0.30 | 0.08 |
|  |  | 0.87 | -0.20 | 0.02 | -0.03 | 0.24 | -0.27 | 0.01 | -0.04 | 0.05 | 0.31 | 0.38 | -0.85 | 0.13 |
| 3BSE dTpdT--dApdA | 237.57 | 0.95 | -0.13 | -0.01 | -0.02 | -0.11 | -0.02 | 0.00 | 0.01 | 0.04 | 0.98 | 0.10 | 0.01 | 0.02 |
|  | 233.73 | -0.10 | 0.96 | 0.00 | 0.07 | 0.02 | -0.08 | 0.00 | 0.00 | 0.03 | -0.10 | 0.98 | 0.03 | 0.00 |
|  | 229.09 | -0.01 | 0.01 | 0.94 | 0.00 | 0.02 | 0.00 | -0.25 | -0.02 | 0.00 | 0.93 | 0.02 | 0.31 | 0.02 |
|  | 226.27 | -0.01 | -0.06 | 0.00 | 0.94 | 0.00 | 0.00 | 0.01 | -0.27 | 0.00 | 0.92 | 0.02 | 0.32 | 0.02 |
|  | 193.23 | -0.01 | 0.00 | 0.54 | 0.00 | -0.02 | 0.00 | 0.80 | 0.01 | 0.05 | -0.37 | 0.02 | 0.90 | -0.02 |
|  | 191.59 | 0.00 | 0.08 | 0.00 | 0.55 | 0.00 | 0.04 | -0.01 | 0.78 | 0.05 | 0.00 | -0.41 | 0.01 | 0.88 |
|  | 188.03 | 0.08 | 0.27 | 0.01 | 0.01 | 0.62 | 0.05 | -0.01 | 0.00 | 0.01 | 0.96 | 0.01 | 0.12 | 0.02 |
|  | 186.87 | 0.06 | 0.01 | -0.02 | -0.02 | 0.59 | -0.07 | -0.04 | 0.01 | 0.03 | 0.89 | 0.21 | 0.13 | 0.05 |
|  |  | -0.01 | 0.00 | 0.04 | 0.01 | -0.29 | 0.00 | -0.02 | 0.02 | 0.03 | 0.34 | -0.01 | 0.05 | -0.01 |
|  | 185.89 | 0.12 | 0.41 | -0.01 | 0.02 | 0.47 | -0.08 | -0.03 | 0.01 | 0.03 | 0.91 | -0.23 | 0.08 | 0.02 |
|  |  | -0.17 | 0.28 | -0.04 | 0.00 | 0.32 | 0.72 | 0.00 | 0.02 | 0.02 | 0.32 | 0.85 | 0.02 | 0.12 |
|  | 184.98 | 0.03 | 0.80 | 0.00 | 0.04 | -0.28 | 0.04 | -0.03 | 0.00 | 0.26 | 0.98 | -0.03 | 0.03 | 0.11 |
| 3BSE dApdTp dTpdG | 240.60 | -0.11 | 0.95 | -0.01 | -0.02 | -0.02 | -0.02 | 0.00 | -0.01 | 0.05 | 0.08 | 0.98 | 0.03 | -0.01 |
|  | 234.53 | 0.96 | -0.08 | 0.02 | 0.00 | -0.03 | 0.02 | 0.03 | 0.00 | 0.04 | 0.98 | -0.06 | -0.04 | 0.03 |
|  | 232.87 | 0.00 | 0.03 | 0.01 | 0.82 | 0.01 | -0.01 | -0.01 | -0.17 | 0.00 | 0.03 | 0.91 | 0.03 | 0.34 |
|  | 226.80 | -0.02 | -0.02 | 0.79 | 0.01 | 0.01 | 0.01 | -0.30 | 0.00 | 0.01 | 0.90 | 0.01 | 0.32 | 0.02 |
|  | 194.66 | -0.01 | 0.10 | 0.00 | 0.41 | 0.00 | -0.01 | 0.00 | 0.69 | 0.10 | 0.00 | -0.47 | 0.00 | 0.84 |
|  | 190.65 | 0.10 | -0.02 | 0.48 | -0.01 | 0.04 | 0.00 | 0.57 | 0.00 | 0.16 | -0.51 | 0.03 | 0.81 | -0.03 |
|  | 188.88 | 0.96 | 0.08 | 0.02 | -0.01 | 0.09 | 0.09 | 0.02 | -0.01 | 0.03 | 0.00 | 0.96 | 0.17 | 0.01 |
|  | 183.95 | 0.93 | -0.06 | -0.05 | 0.00 | -0.04 | 0.02 | -0.03 | -0.01 | 0.08 | 0.03 | -0.18 | 0.56 | 0.02 |
|  | 182.17 | 0.07 | 0.88 | 0.00 | -0.01 | -0.03 | -0.03 | 0.02 | -0.01 | 0.19 | 0.90 | -0.03 | 0.05 | 0.27 |
|  |  | -0.12 | 0.26 | 0.00 | 0.00 | 0.00 | 0.01 | 0.00 | 0.00 | 0.91 | 0.22 | -0.01 | -0.02 | 0.09 |
|  | 176.83 | -0.08 | 0.02 | 0.01 | 0.01 | 0.01 | 0.88 | -0.02 | 0.05 | 0.18 | 0.13 | 0.97 | 0.03 | 0.07 |
|  |  | 0.07 | -0.01 | 0.02 | 0.00 | -0.18 | -0.18 | 0.03 | -0.01 | 0.92 | 0.97 | -0.13 | -0.02 | 0.00 |

| (c) |  | dTpdT--dApdA | | | | dApdTpdTpdG | | | | |
|---|---|---|---|---|---|---|---|---|---|---|
|  | λ (nm) | h-orbital | | e-orbital | | λ (nm) | h-orbital | | e-orbital | |
|  |  | (A1)$^2$ | (A2)$^2$ | (A1)$^2$ | (A2)$^2$ |  | (G)$^2$ | (A)$^2$ | (G)$^2$ | (A)$^2$ |
| Ideal | 242.05 | 0.00 | 0.00 | 0.00 | 0.00 | 242.97 | 0.00 | 0.01 | 0.00 | 0.01 |
|  | 239.09 | 0.00 | 0.00 | 0.00 | 0.00 | 239.62 | 0.00 | 0.00 | 0.00 | 0.00 |
|  | 224.02 | 0.05 | 0.00 | 0.00 | 0.00 | 231.11 | 0.00 | 0.00 | 0.00 | 0.00 |
|  | 223.27 | 0.00 | 0.05 | 0.00 | 0.00 | 230.97 | 0.00 | 0.00 | 0.00 | 0.01 |
|  | 193.00 | 0.00 | 0.00 | 0.00 | 0.00 | 194.60 | 0.00 | 0.00 | 0.03 | 0.01 |
|  | 189.76 | 0.00 | 0.00 | 0.00 | 0.00 | 192.01 | 0.00 | 0.00 | 0.00 | 0.02 |
|  | 188.97 | 0.08 | 0.01 | 0.00 | 0.00 |  | 0.00 | 0.00 | 0.00 | 0.02 |
|  | 186.79 | 0.00 | 0.00 | 0.00 | 0.00 | 190.91 | 0.00 | 0.00 | 0.00 | 0.02 |
|  | 181.87 | 0.05 | 0.02 | 0.00 | 0.00 |  | 0.00 | 0.00 | 0.00 | 0.04 |
|  | 180.96 | 0.01 | 0.45 | 0.02 | 0.01 | 188.45 | 0.00 | 0.01 | 0.00 | 0.19 |
|  |  |  |  |  |  | 186.62 | 0.00 | 0.00 | 0.01 | 0.00 |
|  |  |  |  |  |  | 182.31 | 0.00 | 0.00 | 0.26 | 0.00 |
|  |  |  |  |  |  |  | 0.01 | 0.03 | 0.25 | 0.25 |
| 3BSE | 237.57 | 0.00 | 0.00 | 0.00 | 0.00 | 240.60 | 0.00 | 0.01 | 0.00 | 0.01 |
|  | 233.73 | 0.00 | 0.00 | 0.00 | 0.00 | 234.53 | 0.00 | 0.00 | 0.00 | 0.00 |
|  | 229.09 | 0.03 | 0.00 | 0.01 | 0.00 | 232.87 | 0.00 | 0.00 | 0.00 | 0.00 |
|  | 226.27 | 0.00 | 0.04 | 0.00 | 0.01 | 226.80 | 0.00 | 0.00 | 0.00 | 0.00 |
|  | 193.23 | 0.01 | 0.00 | 0.00 | 0.00 | 194.66 | 0.00 | 0.00 | 0.00 | 0.02 |
|  | 191.59 | 0.00 | 0.01 | 0.00 | 0.00 | 190.65 | 0.00 | 0.00 | 0.01 | 0.00 |
|  | 188.03 | 0.39 | 0.11 | 0.02 | 0.01 | 188.88 | 0.00 | 0.00 | 0.00 | 0.01 |
|  | 186.87 | 0.18 | 0.43 | 0.12 | 0.01 | 183.95 | 0.00 | 0.00 | 0.63 | 0.00 |
|  |  | 0.84 | 0.02 | 0.84 | 0.01 | 182.17 | 0.00 | 0.01 | 0.06 | 0.01 |
|  | 185.89 | 0.18 | 0.37 | 0.06 | 0.02 |  | 0.00 | 0.00 | 0.91 | 0.00 |
|  |  | 0.04 | 0.19 | 0.12 | 0.02 | 176.83 | 0.00 | 0.02 | 0.00 | 0.01 |
|  | 184.98 | 0.00 | 0.01 | 0.00 | 0.00 |  | 0.00 | 0.00 | 0.00 | 0.00 |

**Figure S1.** Various molecular segments extracted from X-ray diffraction determined B-DNA (PDB code 3BSE). The strucutres of (a) dT, (b)dTpdT, (c) dTpdT--dApdA, and (d) dApdTpdTpdG under study are shown.

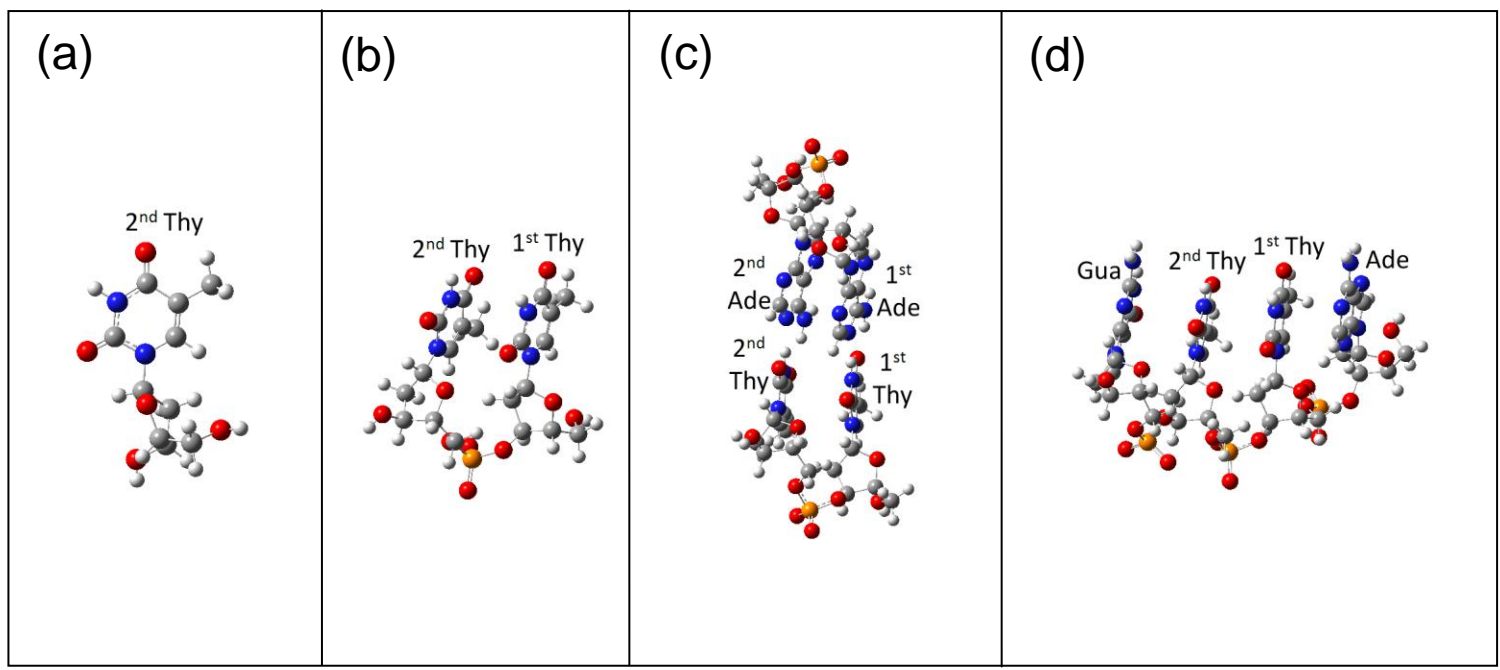

**Figure S2.** Comparison of conformation between ideal and 3BSE dTpdT where the two molecules are superimposed together. The major difference can be found in the backbone bending which in turn results in a small deviation of the base-plane angle. Nonetheless, the important coordinate, *i.e.*, distance between the two thymines, has not deviated strongly from each other. The difference of ground state energy calculated (6-31G(d) DFT/B3LYP) is 0.205eV (3BSE one has lower energy).

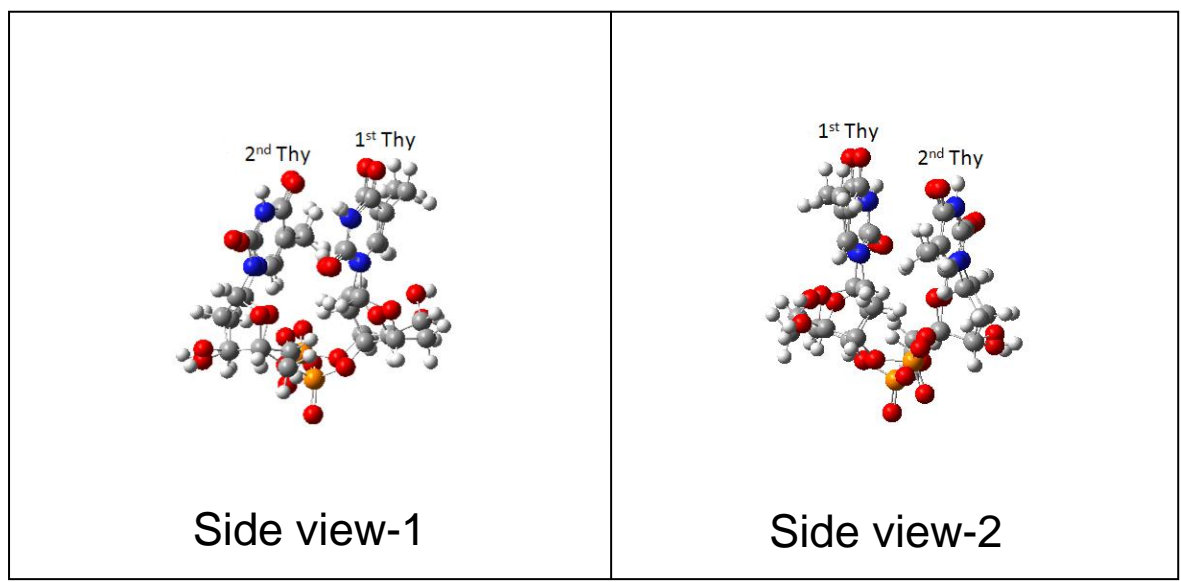

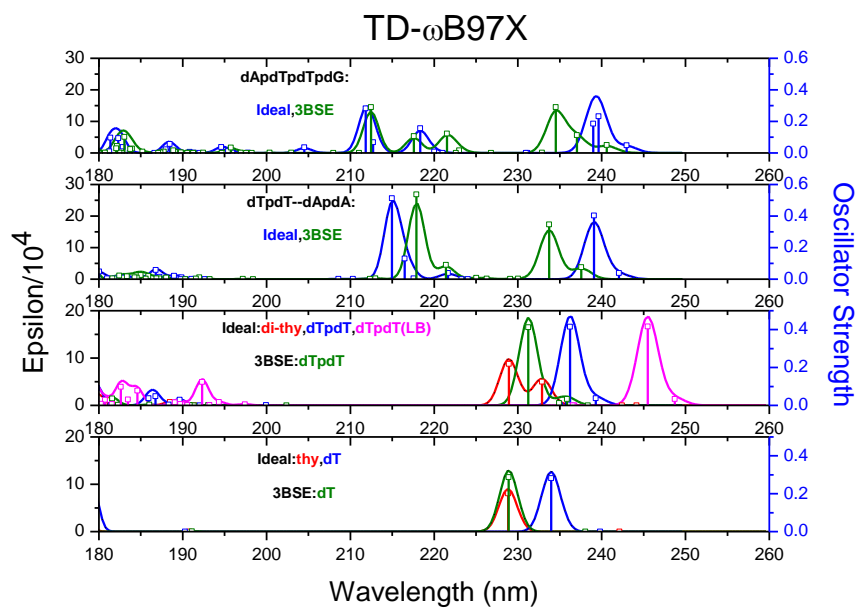

**Figure S3** The calculated absorption spectra of various thymine-comprised ideal/3BSE systems. Each resonance peak is broadened with a Lorentzian with a 0.03eV of Half-Width at Half Height.